\pdfoutput=1

\documentclass[journal]{IEEEtran}

	\usepackage{pifont,bm,multicol,amsfonts,amsmath,color,amssymb,graphicx, epsfig,cite,psfrag,subfigure,algorithm,enumerate,stfloats,algorithm,algorithmic,
epstopdf,balance}
\usepackage{tabularx}
    \usepackage{mathrsfs,pifont,bm,multicol,amsfonts,amsmath,color,amssymb,graphicx, epsfig,cite,psfrag,subfigure,algorithm,enumerate,stfloats,algorithm,algorithmic,
epstopdf,balance,flushend}

\usepackage{booktabs}

\usepackage{xcolor}

\begin{document}
% \textcolor{red}{\cite{FL}} % 在这里，`example` 应替换为你实际的 BibTeX 项名

\title{Channel Estimation and Hybrid Precoding for Massive MIMO-OTFS System  With Doubly Squint}

\author{Mingming~Duan, Pengfei~Zhang,
        Shun Zhang, \emph{Senior Member, IEEE},  \\
         Yao Ge, \emph{Member, IEEE}, 
 Octavia A. Dobre,~\IEEEmembership{Fellow,~IEEE}
     and   
~Chau~Yuen,~\IEEEmembership{Fellow,~IEEE}

\thanks{
Manuscript received 6 November, 2024; revised 19 February, 2025 and 28 March, 2025, accepted 7 April, 2025; date of current version 7 April, 2025.
The work of M. Duan, P. Zhang and S. Zhang was supported by 
the National Key Research and Development
Program of China under Grant 2023YFB2904500
and
the National Natural Science Foundation of China under Grant 62271373.
The work of Yao Ge was supported by the RIE2020 Industry Alignment Fund-Industry Collaboration Projects (IAF-ICP) Funding Initiative, as well as cash and in-kind contribution from the industry partner(s).
\emph{(Corresponding author: Shun Zhang.)}
}

\thanks{Mingming Duan, Pengfei Zhang and Shun Zhang are with the State Key Laboratory of Integrated Services Networks, Xidian University, 710071, China (e-mail: duanmingming@stu.xidian.edu.cn; zhang\_pengfei@stu.xidian.edu.cn; zhangshunsdu@xidian.edu.cn).} 
\thanks{Yao Ge is with Continental-NTU Corporate Laboratory, Nanyang Technological University, Singapore 637553 (e-mail: yao.ge@ntu.edu.sg).} 
\thanks{Octavia A. Dobre is with the Faculty of Engineering and Applied Science, Memorial University, St. Johns, NL A1B 3X5, Canada (e-mail: odobre@mun.ca).}
\thanks{Chau Yuen is with the School of Electrical and Electronics Engineering, Nanyang Technological University, Singapore 639798 (e-mail: chau.yuen@ntu.edu.sg).} 
 }

\maketitle

\begin{abstract}
Orthogonal time frequency space (OTFS) modulation and massive multi-input multi-output (MIMO) are promising technologies for next generation wireless communication systems for their abilities to counteract the issue of high mobility with large Doppler spread and mitigate the channel path attenuation, respectively.
The natural integration of massive MIMO with OTFS in millimeter-wave systems can improve communication data rate and enhance the spectral efficiency.
However, when transmitting wideband signals with large-scale arrays, the beam squint effect may occur, causing discrepancies in beam directions across subcarriers in multi-carrier systems.
%However, beam squint effect can arise in large-scale array when transmitting wideband signals.
%This effect leads to different beam directions among the subcarriers for multi-carrier system.
Moreover, the high-mobility wideband millimeter wave communications
 can induce the Doppler squint effect, leading to different Doppler shifts among the subcarriers.
Both beam squint effect and Doppler squint effect (denoted as doubly squint effect) can degrade communication performance significantly.
In this paper, 
we present an efficient channel estimation and hybrid precoding scheme to address the doubly squint effect in massive MIMO-OTFS systems. 
We first  characterize the wideband channel model and the input-output relationship for massive MIMO-OTFS transmission considering doubly squint effect. 
We then mathematically derive the impact of channel parameters on chirp pilots under the doubly squint effect. Additionally, we develop a peak-index-based channel estimation scheme.
%We then present a channel estimation scheme with chirp-based pilot relying on the peak of discrete Fourier transform-angle sequence signal. 
By leveraging the results from  channel estimation, 
we propose a hybrid precoding method to mitigate the doubly squint effect in downlink transmission scenarios.
Finally, simulation results validate the effectiveness of our proposed scheme and show its superiority over the existing schemes.

% is an effective waveform in high mobility scenarios. In this paper, we focuses on a massive multi-input multi-output system with OTFS waveform adopted.
% To enhance the transmission performance, OTFS combined with broadband massive multi-input multi-output (MIMO) technology has attracted research attention. However, with the increase of antenna array and bandwidth, the beam dispersion effect and Doppler dispersion effect are caused, which degrade the transmission performance. In this paper, we analyze the impact of dual-dispersion channel on massive MIMO OTFS system under the influence of beam dispersion and Doppler dispersion, and propose hybrid precoding to improve the communication rate under dual-dispersion channel. Firstly, a channel estimation model based on chirp sequence is proposed. By using the true delay line, the large-scale array is equivalent to multiple sub-arrays, which reduces the influence of beam dispersion on channel estimation. Then, the hybrid and coding structure was proposed, and the precoding scheme for ISI and ICI was derived. The boundary conditions of ISI and ICI were derived, and the beam dispersion and Doppler dispersion as well as the influence of ISI and ICI were reduced.
% Simulation results show that it has better performance in massive MIMO scenarios with high mobility.
\end{abstract}


\IEEEpeerreviewmaketitle

\section{Introduction}

\IEEEPARstart{F}{utrue} wireless networks are anticipated to achieve high-throughput  multi-user access in high-mobility scenarios.
This demand drives the exploration  of 
 millimeter wave (mmWave) communication, which is capable of utilizing the 30 to 300 GHz frequency range \cite{bg_3}.
 Despite mmWave capability of delivering multigigabit data rate,
its high carrier frequency results in server path loss in comparison to systems operating in the sub-6 GHz frequency range \cite{bg_0,bg_4}.
On the other hand, the reduction in wavelength of mmWave systems facilitates a high density of antennas within compact spaces.
This advantage further accelerates the deployment of massive multi-input multi-output (MIMO) technology
% allows a high density of antennas within a compact spaces. 
%This advantage further accelerate the deployment of massive multi-input multi-output (MIMO) technology
\cite{beam_1}.
In this case, the path loss can be compensated through the use of directional beams and high array gains.
Considering these characteristics, researchers have investigated to exploit the potential of massive MIMO in mmWave systems.
%Considering these characteristics, researchers have investigated to exploit the potential of massive MIMO.
In \cite{massive_MIMO_2}, Zhang \emph{et al.}  proposed beamforming schemes based on the reciprocity of dominant angle of arrival (AoA), where the overhead of beamforming is reduced effectively.
% reciprocity of dominant angle of arrival (AoA) to increase the array gain.
In \cite{massive_MIMO_4}, Almoneer \emph{et al.}  deployed a hardware platform that incorporates different
 variants of zero-forcing precoding to investigate the performance of MIMO transmitters.
{
Moreover, the capacity of massive user access in massive MIMO systems supports
 the implementation of distributed edge computing and federated learning in wireless communications \cite{dis_comp, FL}.
Cai \emph{et al.}  proposed a massive NOMA scheme for uplink and downlink transmissions in cellular IoT communications \cite{massiveNOMA}.
}
Furthermore, 
a low-complexity architecture is considered in hybrid precoding structures
%hybrid precoding structures are considered a low-complexity architecture 
for massive MIMO systems \cite{mmwave_1,mimo_1,mimo_2}.
These structures are cable of reducing the usage  of the radio frequency (RF) chains and can also maintain the full multiplexing gains by using the inherent sparsity of wireless channel \cite{mmwave_1,mimo_1,mimo_2,hybrid_3}.
In \cite{hybrid_2}, Ramadan \emph{et al.} proposed a hybrid precoder design leveraging artificial noise and further investigated the impact of phase shifters (PSs) resolution. 
In \cite{hybrid_4}, Gu \emph{et al.} 
examined the relationship between the spectral efficiency, interleaved structures, and antenna spacing, where an optimal interval for hybrid precoding design was proposed for subconnected architectures.

% and proposed an optimal hybrid precoding design.
%In \cite{hybrid_1}, Hajjaj \emph{et al.} introduced a iterative hybrid precoder to reduce computational complexity.
%% which employs an iterative semidefinite relaxation approach to significantly reduce computational complexity.

On the other hand, in wideband systems equipped with a substantial number of antennas,
 the propagation delay variation among antennas can be comparable to the sampling period, making the time delay across the array non-negligible. 
This phenomenon can further lead to the beam squint effect \cite{squint_16}, resulting  the beam direction to vary with subcarriers for the transmission of multi-carrier signals \cite{MIMO_6,squint_2,squint_5}.
Due to the loss of the array gain along the subcarriers,
 the traditional beamforming algorithms exhibit a significant performance degradation\cite{squint_17}.
Fortunately, the beam squint effect can be compensated for either at the receiver or transmitter. 
Specifically, on the receiver side, the communication performance can be improved primarily by improving channel estimation techniques in the context of beam squint effect.
%In order to fully harness the potential of massive MIMO system, it is crucial to obtain accurate channel state information (CSI) in the presence of beam squint effect.
In \cite{squint_14}, Maity \emph{et al.} proposed a  channel estimation scheme utilizing dictionary-learning based to counteracts beam squint effect.
%that exploits the channel sparsity to improve channel state information estimation.
In \cite{squint_18}, Shi \emph{et al.} 
proposed a frequency-dependent estimation method to minimize  the beam squint effect, where a frequency dependent dictionary was established for each subcarrier. 
%addressed the challenge of channel estimation in mmWave systems with large antenna arrays by proposing a spatial spectrum-based scheme for wideband channel estimation that accounts for the beam squint effect
%Therefore, it is necessary to compensate the gain loss caused by beam squint effect to improve capacity \cite{squint_2}, \cite{squint_5}.
%Furthermore, the array gain loss caused by beam squint effect can be mitigated by the transmitters.
At the transmitter side, extensive studies have been dedicated to mitigate the reduction in array gain  resulting from the beam squint effect.
%On the other hand, extensive research has been conducted on mitigating the beam squint effect at the transmitter.
%it is necessary for the transmitter to mitigate the array gain loss caused by beam squint effect.
%There are many studies on reducing the effects of beam squint.
In \cite{squint_9}, Cai \emph{et al.} developed a codebook design that maximizes the array gain across all subcarriers to mitigate  the beam squint effect.
In \cite{squint_19}, Zhang \emph{et al.} proposed an  alternative iterative algorithm for both fully connected and subconnected array architectures to alleviate the effect of beam squint and enhance the average rate.
In \cite{squint_4}, Dai \emph{et al.} introduced a delay-phase precoding (DPP) to reduce the loss of array gain experienced over the entire bandwidth.

Doppler caused by mobility between transmitter and receiver is another notable factor in wireless communication. 
It results in a time-varying channel with significant changes in both amplitude and phase over time \cite{high_mob_1}. 
Therefore, the performance of the modulation technique employed in 4G and 5G neworks, i.e., orthogonal frequency division multiplexing (OFDM), may deteriorate due to the degradation of orthogonality among subcarriers.
%The current orthogonal frequency division multiplexing (OFDM) system has the advantage of multipath fading, but the ICI caused by Doppler will lead to severe performance degradation
Fortunately, Hadani \emph{et al.} proposed orthogonal time frequency space (OTFS) modulation \cite{OTFS_1}. which can
exploit the diversity of both channel delays and Doppler spreads.
OTFS demonstrates a distinct advantage over OFDM in high mobility communications with large Doppler spread.
By integrating OTFS with massive MIMO systems, 
the channel can be further scheduled in a three-dimensional (3-D) Doppler-delay-angle domain for performance enhancement with pertinent resource allocation schemes \cite{OTFS_4,OTFS_8}.
%the communication system leverages the sparsity inherent in the three-dimensional (3-D) Doppler-delay-angle domain to enable more intricate resource scheduling strategies  
%the communication performance can be further adapted to the future wireless network
In \cite{OTFS_9}, Pandey \emph{et al.} presented 
a multi-user precoder and detector for massive MIMO-OTFS system, where the complexity was reduced by independently  performing detection for each information symbol.
In \cite{MIMO_OTFS_1}, Shao \emph{et al.} introduced a 3-D non-orthogonal multiple access scheme to exploit  the delay-Doppler domain resources with the overlapped angle signature.
In \cite{MIMO_OTFS}, Shen \emph{et al.} 
adapted massive MIMO-OTFS for satellite communications to address the issues arising from high mobility and massive access in satellite-based communications.
In \cite{OTFS_7}, Li \emph{et al.} constructed a path division multiple access (PDMA) scheme, which fully utilizes the scattering path resolution by assigning angle-domain resources to different users.
%maximized the use of scattering path resolution by allocating angle-domain resources to different users.

However, in the case of wideband signal transmissions, the high mobility can cause Doppler squint effect (DSE), resulting in frequency-dependent Doppler shifts.
This effect can particularly disrupt sparsity of the channel, leading to inaccuracy in channel estimation and data detection because of modeling error.
Most of the current research on OTFS assumes the channel exhibits substantial sparsity in the DD domain, where the DSE is not considered.
Nonetheless, it has been proved that the performance of OTFS systems can be severely compromised if DSE is disregarded \cite{squint_3}.
%Hence, without considering Doppler shift effects (DSE), OTFS modulation performance can experience significant degradation \cite{squint_3}.
%In OTFS system, DSE is accumulated over a longer duration, which prevents OTFS from taking advantage in high mobility scenarios.
%the phase deviation across subcarriers can be accumulated in one OTFS symbol.
%it has been proved 
%In , Wang \emph{et al.} proved 
%that the performance of OTFS modulation can be severely degraded by Doppler squint effect \cite{squint_3}.
For the improvement of OTFS communication performance in high-mobility wideband scenarios, it is imperative to account for the impact of DSE.
In \cite{squint_11}, Wang \emph{et al.} proposed a  parameter estimation scheme based on sparse Bayesian learning, thereby enhancing the accuracy of channel estimation for OTFS systems with DSE.
In \cite{squint_20}, Wang \emph{et al.} emphasized the importance of accounting for DSE in OTFS systems and introduced a low-complexity receiver design incorporating linear equalization, which improves performance of OTFS in the presence of DSE.
%Therefore, it is important to consider Doppler squint effect in channel estimation and precoding.

Future communication systems with high-mobility will likely encounter both beam squint and Doppler squint effects simultaneously.
Therefore, it is necessary and important to develop novel communication schemes that can effectively handle both beam squint effect and Doppler squint effect.
% addressed Doppler squint and beam squint effects independently without considering both of them.
In \cite{squint12}, Liao \emph{et al.} introduced a channel estimation scheme to address  performance degradation resulting from Doppler squint effect and  beam squint effect for massive MIMO-OFDM systems.
However, 
the scheme proposed in \cite{squint12} can not be directly applied for massive MIMO-OTFS systems subject  to high mobility challenges. 
%the use of OFDM modulation results in reduced performance under high-mobility conditions.
%OTFS modulation is capable of addressing the challenges of high-mobility communication. 
In addition,
the existing studies in the field of massive MIMO-OTFS systems are mainly focused on addressing either beam squint effect or Doppler squint effect separately, rather than addressing both simultaneously.
In order to enhance the performance of massive MIMO-OTFS systems under future high mobility and wideband transmissions, we introduce an efficient channel estimation and hybrid precoding scheme in this work based on the doubly-squint effect, i.e., beam squint and Doppler squint channel conditions.
The main contributions of our study can be summarized as follows:
\begin{itemize}
%This paper describes the uplink estimation and downlink transmission process.
\item 
We introduce a massive MIMO-OTFS system model
based on the doubly-squint effect.
We derive the input-output relationship in the DD domain with the proposed hybrid precoding scheme.
Meanwhile,
we discuss and analyze the specific effect of Doppler squint and beam squint in wideband systems compared to narrowband systems.

\item 
We propose a channel estimation structure based on chirp pilot and develop a parameter estimation method  considering the doubly squint effect.
%Specifically, we proposed a TTD
Specifically, we mathematically derive the optimal design of true time delay (TTD) and  phase shifter (PS) for chirp pilots under doubly squint effect. 
Additionally, we establish the relationship between peak index and channel parameters. Then we propose a peak-based channel estimation scheme.

%we derive the optimal settings for true time delay (TTD) and PS to maximize the array gain.
%We further
%propose a channel parameter estimation method using the peak of DFT-angle sequence for up-chirp pilot and down-chirp pilot.
%The 2-D Jocabsen estimator is used to increase the accuracy of estimation. 
%joint estimation of DFT-angle sequences using up-chirp and down-chirp pilots.
%In contrast to conventional multi-carrier pilot designs, we propose a channel estimation with chirp pilot, which is a linear frequency modulated signal with a chirp cyclic prefix (CCP). 
%We design the  true time delay (TTD) and phase shifter (PS) configurations at the BS receiver for our estimation, and proposed a low-complexity parameter estimation scheme based on the peak of the demodulated chirp pilot.

\item 
We propose a hybrid precoding method (including digital and analog precoding) to address the doubly squint effect. Specifically, the analog precoding addresses the beam squint effect, while the digital precoding mitigates the Doppler squint effect.

%with PDMA scheme and analyze the compensation for doubly squint effect and.
%We propose a hybrid precoding scheme to mitigate the impact of doubly squint effect.
%Moreover, 
%the impact of inter-carrier interference (ICI) and inter-symbol interference (ISI) can be eliminated in our method, respectively. 

\item
%The proposed channel estimation and hybrid precoding method take both beam squint effect and Doppler squint effect into consideration. 
Simulation results demonstrate that 
the proposed channel estimation and precoding method can achieve stronger performance than the existing methods and exhibit robustness to wideband massive MIMO-OTFS system.
\end{itemize}

The remainder of this paper is organized as follows. 
Section II introduces the system model of massive MIMO-OTFS, where the input-output relationship considering the doubly squint effect is presented. 
In Section III, we introduce the chirp pilot into the wideband massive MIMO system and propose a novel channel parameter estimation framework.
Section IV develops the hybrid precoding schemes, which addresses the beam squint and Doppler squint effect through analog precoding and digital precoding, respectively. 
Simulation results are presented in Section V  and the conclusions are delineated in Section VI, respectively

\emph{Notation:}
%Denote lowercase (uppercase) boldface as vector. 
$(\cdot)^*$ represents conjugate.
The real part and the imaginary part of $x$ are expressed as $\Re(x) $ and $\Im(x) $, respectively.
% $\lfloor s \rfloor$ returns the largest integer less than or equal to $x$ and $\lceil x \rceil$ returns the smallest integer greater than or equal to $s$.
 Let $\left\| x \right\|$ denote the Euclidean norm. $\mathbb{E}(\cdot)$ represents the expectation. $| \cdot |$ is the absolute value.
$(*)$ denotes convolution.
%$\odot$ denotes the Hadamard product and $\otimes$ denotes Kronecker product.
 $[\cdot]_N$ is the MOD $N$ processing. 
 Finally, $\delta(\cdot)$ represents the Dirac delta function.

%   \begin{figure}[htbp]
%  \centering
%  \includegraphics[width=75mm]{scenario.png}
%  \caption{Massive MIMO-OTFS system model.}
%  \label{scenario}
% \end{figure}

\begin{figure*}[!htbp]
 \centering
 \includegraphics[width=170mm]{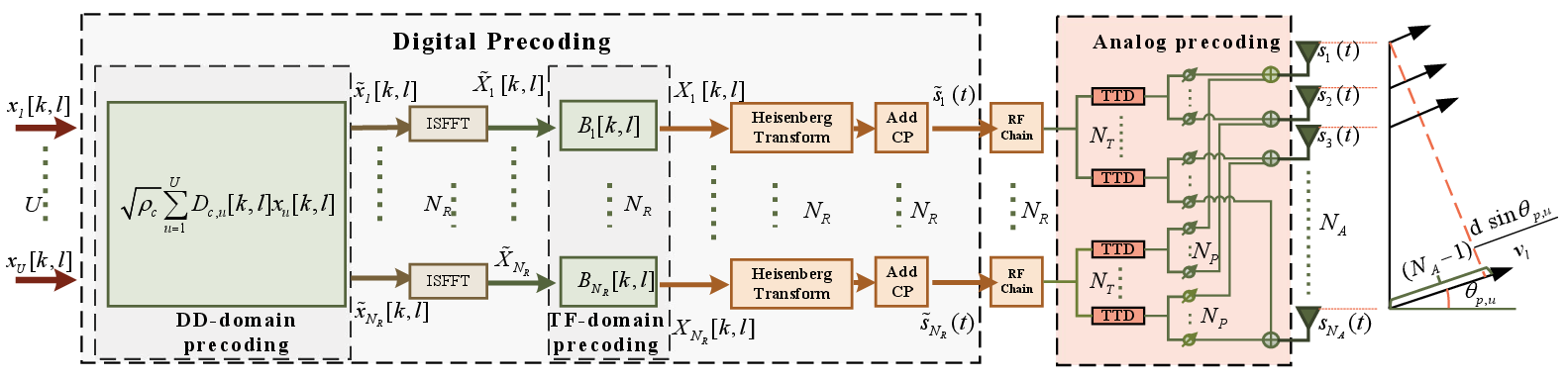}
 \caption{Massive MIMO-OTFS system with Hybrid structure.}
 \label{Massive MIMO-OTFS Hybrid_structure}
\end{figure*}

\section{System Model}

We examine a wideband mmWave massive MIMO-OTFS system, which includes a base station (BS) and $U$ single-antenna users. It is assumed that there are $P$ scattering paths between the BS and each user. The BS is furnished with a uniform linear array (ULA) that comprises $N_{A}$ antennas and $N_{R}$ RF chains. As shown in {
\figurename{ \ref{Massive MIMO-OTFS Hybrid_structure}} }, each RF chain is connected to $N_{T}$ independent TTDs, i.e., the TTD set connected to the $c$-th RF chain is defined as $ \mathbf{T}_{c} = \{\mathbf{T}_{c,1}, \mathbf{T}_{c,2}, \ldots, \mathbf{T}_{c,N_{T}}\} $
with
$\mathbf{T}_{c} \cap \mathbf{T}_{c^{\prime}} = \emptyset $
when $c \neq c'$, where $\mathbf{T}_{c,d}$ denotes the $d$-th TTD in $\mathbf{T}_{c}$, $d=1,2,\ldots, N_{T}$ and $c=1,2,\ldots, N_{R}$. 
Each TTD is then connected to $N_{P} = N_{A} / N_{T}$ PSs, i.e., the PS set connected to $\mathbf{T}_{c, d}$ is defined as
$ \mathbf{P}_{c, d} = \{\mathbf{P}_{c, d, 1}, \mathbf{P}_{c, d, 2}, \ldots, \mathbf{P}_{c, d, N_{P}}\} $
with
$ \cup_{c=1}^{N_{R}} \cup_{d=1}^{N_{T}} \mathbf{P}_{c, d} = \Omega $
and
$ \mathbf{P}_{c, d} \cap \mathbf{P}_{c^{\prime}, d^{\prime}} = \emptyset $
either when $c \neq c'$ or $d \neq d'$, where $\mathbf{P}_{c, d, s}$ denotes the $s$-th PS in $\mathbf{P}_{c, d}$, $s=1,2,\ldots, N_{P}$ and $|\Omega| = N_{A} N_{R}$. Finally, each antenna is connected to $N_{R}$ PSs, i.e., the PS set connected to the $a$-th antenna is defined as
$ \mathbf{P}_{a}^{\prime} = \{\mathbf{P}_{a, 1}^{\prime}, P_{a, 2}^{\prime}, \ldots, \mathbf{P}_{a, N_{R}}^{\prime}\} $
with
$ \cup_{a=1}^{N_{A}} \mathbf{P}_{a}^{\prime} = \Omega $
and
$ \mathbf{P}_{a}^{\prime} \cap \mathbf{P}_{a^{\prime}}^{\prime} = \emptyset $
when $a \neq a'$. Here, $\mathbf{P}_{a, c}^{\prime} = \mathbf{P}_{c, d, s}$ represents the same PS for $a = (d-1) N_{P} + s$.

\subsection{OTFS Transmitter}

We assume that the BS can transmit independent OTFS symbols to $U$ users simultaneously. As shown in {
\figurename{ \ref{Massive MIMO-OTFS Hybrid_structure}}}, the $U$ OTFS symbols are first processed by digital precoding, which includes DD-domain precoding and TF-domain precoding. Let $\mathbf{x}_u \in \mathbb{C}^{N \times M}$ denote OTFS symbol transmitted to the $u$-th user, $u = 1,2,\ldots, U$. Then, the OTFS symbols are precoded in DD domain to generate $N_{R}$ independent OTFS symbols for $N_{R}$ RF chains. After DD-domain digital precoding, the OTFS symbol related to the $c$-th RF chain is given by
\begin{align}\label{DD comp}
\tilde {x} _{c}[k,\ell]= 
%\sqrt{\rho_{a}}
{\sqrt {\rho _{c}}}
\sum_{u=1}^{U}{D}_{c,u}[k,\ell]{ x}_u[k,\ell],
%s=&0,1,\dots,S,
c=1,\dots , N_R,
\end{align}
where $x_{u}[k,\ell] $ is the element of $\mathbf {x}_u$ in the $k$-th row and $\ell$-th column,
 $\sqrt{\rho _{c}}$ is the power factor and ${D}_{c,u}[k,\ell]$ is the DD-domain digital precoding element.
With the help of the inverse symblectic finite Fourier transform (ISFFT), we can derive
the TF domain symbol $\tilde X_{c}[n,m] $, i.e.,
\begin{align}\label{Xp tf domin}
\tilde X_{c}[n,m] = \frac{1}{\sqrt{NM}} 
\sum_{k = 0}^{N-1}\sum_{\ell = 0}^{M-1}\tilde x_{c}[k,\ell]
e^{j2 \pi (\frac{nk}{N}-\frac{m\ell}{M} )}, \notag \\
n=0,...,N-1, m=0,...,M-1.
\end{align}
Subsequently, TF-domain digital precoding is implemented to mitigate the subcarrier-dependent and time-dependent phase shifts, expressed as
\begin{align}\label{TF comp}
 X_{c}[n,m] =  B_{c}[n,m] \tilde X_{c}[n,m] ,
\end{align}
where $B_{c}[n,m]$ is the TF-domain digital precoding element.
We then obtain the continuous baseband time-domain signal $\tilde s_{c}(t)$ using the Heisenberg transform given by
\begin{align} 
\tilde s_{c}(t)  = &\sum_{n = 0}^{N-1}\sum_{m  = 0}^{M-1} X_{c}[n,m]g_{tx}(t-nT)
e^{j2\pi m\Delta f(t-nT)}  , 
\end{align} 
where $g_{tx}(t)$, $T$ and $\Delta f$ denote the transmit pulse, the symbol duration and
subcarrier interval, respectively.
In order to counteract the effect of multipath,
a cyclic prefix (CP) is added with the length exceeding the maximum delay spread.

As shown in {
\figurename{ \ref{Massive MIMO-OTFS Hybrid_structure}}}, the final time-domain signal $\mathbf{s}(t) = \left\{s_{1}(t), s_{2}(t), \ldots, s_{N_{A}}(t)\right\}^{T}$ can be derived by applying analog precoding to the signals from $N_{R}$ RF chains. For the signal from the $c$-th RF chain, it is forwarded to the $a$-th antenna via the TTD $\mathbf{T}_{c,d}$ and the PS $\mathbf{P}_{a,c}^{\prime}$ (i.e., $\mathbf{P}_{c,d,s}$). Therefore, the time-domain response from the $c$-th RF chain to the $a$-th antenna (i.e., the equivalent time-domain response introduced by $\mathbf{T}_{c, d}$ and $\mathbf{P}_{c, d, s}$) can be expressed as
% As shown in \figurename{ \ref{Massive MIMO-OTFS Hybrid_structure}}, 
% the final time-domain signal
% the $c$-th RF chain is connected to the $a$-th antenna through TTD $\mathbf{T}_{c,d}$ and PS $\chi _{c,d,s}$. 
% Therefore, the time-domain response introduced by the conjunction of these TTD and PS can be expressed as
\begin{align}\label{ttd ps resp}
  %{ s _{n_{RF} , \kappa,\varepsilon}}(t) = 
%{\tilde s_p}(t){e^{ - j2\pi {\Psi _{\kappa ,\varepsilon ,{n_{RF}}}}}} \otimes \delta (t - {t_{\kappa ,{n_{RF}}}}),
\bar{\rm  h}_{c,d,s}(t) =
\delta
(t-{t_{c ,d}}) {e^{  j2\pi {\Psi _{c ,d ,s}}}},
% \otimes \delta (t - {t_{\kappa ,{n_{RF}}}})
\end{align}
where $t_{c,d}$ is the delay of the $\mathbf{T}_{c,d}$ and $e^{j2\pi \Psi _{c ,d ,s}}$ expresses the phase shift of $\mathbf{P} _{c,d,s}$.
Thus, the $a$-th element of $\mathbf{s}(t)$
can be derived as
% Then, the corresponding baseband signal $s_{a}(t)$ at the $a$-th antenna  is derived from $N_R$ RF chains as
%
% the transmit signal at the $((b-1) N_P+ c)$-th antenna can be expressed as
\begin{align}\label{trans signal at r}
 s_{a}(t)=\sum_{c=1}^{N_{R}}
\tilde s_c(t-{t_{c ,d}}) 
{e^{ j2\pi {\Psi _{c ,d ,s}}}}, a=(d-1)N_P +s, \notag \\
d=1,2,\dots,N_T, s=1,2,\dots,N_P.
 %a =1,2,...N_A.
%{ s _{n_{RF} , \kappa,\varepsilon}}(t)
\end{align}
% where $a$ can be derived by $a=(d-1)N_T + s-1$.
%To simplify notation, we define $q=(b-1) N_P + c$. Hence, the transmit signal $s_{b,c}(t)$ can be represented as $s_{q}(t)$.
%where $r=b N_P + c$ is the index of antenna, that is, $\kappa = \lfloor r / N_P \rfloor$ and $\varepsilon = (r \mod N_P)$.

\subsection{Massive MIMO-OTFS Channel Model}

In the majority of studies related to massive MIMO, channel parameters are assumed to remain constant within a snapshot. 
% The variations in angle and Doppler are minute within the snapshot and are bandwidth-insensitive. 
% Thus the angle and Doppler can be regarded as constant within the snapshot.
However, in wideband systems, the delay variation, which leads to the Doppler squint effect,  cannot be ignored within a snapshot \cite{squint_3}.
 \footnote{ 
According to the 3rd Generation Partnership Project (3GPP) Technical Report 38.901 \cite{38901}, it is acceptable for the channel parameters to be updated within a distance variation less than 1 meter. 
In this paper, the distance of user moving within an OTFS symbol period and the channel estimation period can be much less than 1 meter.
For the sake of simplification, Doppler and angle are still considered constant within a snapshot due to their insensitivity to bandwidth.
}
In this work, we consider a wideband channel model, where the TF-domain channel response between the $a$-th antenna of BS and the $u$-th user can be expressed as 
\begin{align}\label{H_tf}
H_{u,a}(t,f) = &\sum_{p=1}^{P} \tilde{\alpha }_{p,u}e^{-j2\pi( f_c + f) \tau_{p,u,a} } e^{j2\pi \frac{\nu_{p,u}}{f_c}(f_c+f)t}
\notag \\
=&\sum_{p=1}^{P} \tilde{\alpha }_{p,u}e^{-j2\pi( f_c + f) \left( \tau_{p,u,a} - \frac{\nu_{p,u}}{f_c}t \right )},
\end{align}
where $ \tilde{\alpha }_{p,u}$, $\nu_{p,u}=\frac{v_{p,u}f_c}{v_l}$ and $v_{p,u}$ denote the complex channel gain, Doppler shift and velocity of the $p$-th path with $u$-th user, respectively.
% {\color{blue} \footnote{ \color{blue}
% As in \cite{squint_3}, Doppler squint effect is manifested as time-varying delay in the time domain response.
% In traditional massive-MIMO studies, the assumption that channel parameters remain static within a snapshot is common. 
% However, in wideband conditions, delay variations within the snapshot are significant. 
% This arises because broadband systems have high delay resolution, making them more sensitive to such changes.
% }}
 $v_l$  and $f_c$ denote the velocity of light and the carrier frequency. 
$\tau_{p,u,a}$ is the propagation delay corresponding to $a$-th BS antenna of the $p$-th path with $u$-th user and can be written as
\begin{align}
 \tau_{p,u,a}=\tau_{p,u}+(a-1) \frac{ {\rm d} \sin(\theta_{p,u})
}{f_c \lambda_c} ,
\end{align}
where $\tau_{p, u}$ and $\theta_{p, u}$ denote the propagation delay and the physical direction the $p$-th path with the $u$-th user. $\rm{d}$ and $\lambda_{c}$ denote the antenna spacing and carrier wavelength, respectively. 
In this work, we set $\rm{d} = \frac{\lambda_{c}}{2}$ and define the spatial direction as $\psi_{p, u} = \frac{\rm{d} \sin(\theta_{p, u})}{\lambda_{c}}$. 
Subsequently, the DD-domain channel response $h_{u, a}(\tau, \nu)$ can be derived by the 2D symplectic finite Fourier transform (SFFT) of $H_{u, a}(t, f)$ \cite{squint_3} as (9), at the top of the next page, where $\alpha_{p, u} = \tilde{\alpha}_{p, u} e^{-j 2\pi f_{c} \tau_{p, u}}$ is the equivalent path gain.

\subsection{OTFS Receiver}

\begin{figure*}[!htbp]
\begin{align}\label{h_tau_v}
 h_{u,a}(\tau ,\nu)  =
\left\{ \begin{array}{ll}
 \sum_{p = 1}^{P} \alpha _{p,u}  e^{-j2\pi (a-1){\psi _{p,u}}}  \left | \frac{f_c}{\nu_{p,u}}  \right | 
 e^{2\pi \frac{f_c}{\nu_{p,u}} \left (\tau -\tau _{p,u}-(a-1)\frac{\psi _{p,u}}{f_c} \right )(\nu-\nu_{p,u}) } 
,&
 \nu_{p,u}\ne 0  \hfill \\
  \sum_{p = 1}^{P} \alpha _{p,u}
   e^{-j2\pi (a-1){\psi _{p,u}}}
  \delta \left(\tau -\tau _{p,u}-(a-1)\frac{\psi _{p,u}}{f_c} \right) \delta (\nu) ,&\nu_{p,u} =0 \hfill \\ 
\end{array}  \right. ,
\end{align}
\hrulefill
\end{figure*}
At the $u$-th user, the CP is first removed and the received time-domain signal can be formulated as
\begin{align}\label{recv_ys}
\tilde y_{u}(t) \!=&\! \! 
\sum_{a=1}^{N_A}\! \! 
%\sum_{n_{RF}=1}^{N_{RF}}
 \iint \!  \! \!  h_{u,a}(\tau \!,\nu)e^{j2\pi \nu(t-\tau)}\!
 s_{a} \! (  t\! - \! \tau)d\tau d\nu
\! + \!w_u \! (t) ,
\end{align}
where $w_u(t)$ is the additional white Gaussian noise (AWGN) at the $u$-th user with variance $\sigma^2$.
%
%
%Substituting \eqref{trans signal at r} into \eqref{recv_ys},
%the signal  $y_s(t)$ can be rewritten as
%\begin{align}
%y_{s}(t) =&
%%\sum_{r=1}^{N_r}
%\sum_{a=1}^{N_{R}}
% \iint \tilde h_{s,a}(\tau,\nu)e^{j2\pi \nu(t-\tau)}
%\tilde s_{a}(t-\tau)d\tau d\nu 
%%\notag \\
%%&
%+w_s(t),
%\end{align}
%where $\tilde h_{s,a}(\tau,\nu)$ is the equivalent DD domain response from BS's $a$-th RF chain to the $s$-th user and can be expressed as
%
%\begin{align}\label{h_tau_v2}
%\tilde h_{s,a}(\tau,\nu) =& \sum_{p = 1}^{P}
%\sum_{q=1}^{N_r}
% \alpha _{p,s}\left | \frac{f_c}{\nu_{p,s}}  \right | 
%e^{-j2\pi f_c({t_{a ,b}}+(q-1)\frac{\psi _{p,s}}{f_c} )} \notag \\
%&\times e^{2\pi \frac{f_c}{\nu_{p,s}}(\tau - {t_{a ,b}} - \tau _{p,s}-(q-1)\frac{\psi _{p,s}}{f_c} )(\nu-\nu_{p,s}) } \notag \\
%&\times {e^{  j2\pi {\Psi _{a ,b ,c}}}}.
%%for v_{p,i}\ne 0 .
%\end{align}
By using Wigner transform, the TF domain signal $Y_{u}(t,f)$ can be derived as
 %received by the $p$-th user, as shown
 %in \eqref{signal from dd}
%is transformed to the  using Wigner transform, i.e.
\begin{align}
  Y_{u}(t,f)=&\int g^*_{rx}(t'-t) \tilde y _{u}(t')e^{-j2\pi ft'}dt' ,
\end{align}
where $g_{rx}(t)$ is the receive pulse.
% For simplicity, both $g_{tx}(t)$ and $g_{rx}(t)$ are assumed to be rectangular pulses.
We sample the output $  Y_{u}(t,f)$ as ${Y_u}[n,m]=Y_{u}(t,f)|_{t=nT,f=m\Delta f}$.
  Then, the TF-domain symbol ${Y_u}[n,m]$ can be represented as
\begin{align}\label{Yp}
{Y_u}[n,m] =& 
%\sum\limits_{i = 1}^{{L_p}}
 {\sum\limits_{n' = 0}^{N - 1}
  {\sum\limits_{m' = 0}^{M - 1} 
  \sum_{c=1}^{N_{R}}
%\sum_{r=0}^{N_r-1}
{ H_{n,m}^{u,c}[n',m']{X_c}[n',m']} } } 
%\notag \\
%&
+ {W_u}[n,m], \notag \\
n=&0,1,\dots ,N-1,m=0,1,\dots,M-1,
\end{align}
where $W_u[n,m]$ is transformed and sampled from $w_u(t)$ by the Wigner transform.
 ${ { H}_{n,m}^{u,c}}[n',m']$ is the channel matrix expressed as
 \eqref{Hp}, at the top of the next page,
\begin{figure*}[!htbp]
\vspace{-0.3cm}
\begin{equation}\label{Hp}
  H^{u,c}_{n,m}[n',m'] = 
 %\sum_{q=1}^{N_r}
 \iint \tilde h_{u,c}(\tau ,\nu) A_{g_{rx},g_{tx}}((n-n')T-\tau ,(m-m')\Delta f-\nu)  e^{j2\pi (\nu-m \Delta f)(\tau +n'T)} d\tau d\nu,
\end{equation}
\vspace{-0.3cm}
\begin{align}\label{h_tau_v2}
\tilde h_{u,c}(\tau ,\nu)  = \!
\left\{  \!
\begin{array}{ll} 
 \sum_{p = 1}^{P} \! \sum_{a = 1}^{N_A} 
 \!
 \alpha _{p,u} 
 \!
  e^{-j2\pi f_c \left(t_{c,d} + ( \! a \! -\! 1 \!)\frac{\psi _{p,u}}{f_c} \right) } 
  \!
 \left | \frac{f_c}{\nu_{p,u}}  \right | 
 e^{2\pi \frac{f_c}{\nu_{p,u}}\left( \! \tau \! - \! t_{c,d} \! - \! \tau _{p,u} \! - \! ( \! a \! - \! 1 \!)\frac{\psi _{p,u}}{f_c} \right)(\nu \! - \! \nu_{p,u}) } 
% e^{-j2\pi f_c \left({t_{c ,d}}+(a-1)\frac{\psi _{p,s}}{f_c} \right)} 
{e^{  j2\pi {\Psi _{a ,b ,c}}}}, \!&
 \nu_{p,u} \! \ne \! 0 \! \hfill \\
  \sum_{p = 1}^{P} \! \sum_{a = 1}^{N_A} 
 \!
 \alpha _{p,u} 
 \!
  e^{-j2\pi f_c \left(t_{c,d} + ( \! a \! -\! 1 \!)\frac{\psi _{p,u}}{f_c} \right) } 
  \!
\delta \left(\tau \! -t_{c,d} \! -\tau _{p,u} \! -( \! a \! -1)\frac{\psi _{p,u}}{f_c} \right)\delta (\nu)
{e^{  j2\pi {\Psi _{a ,b ,c}}}}
,&\nu_{p,u} =0 \hfill \\ 
\end{array}  \right. \! ,
\end{align}
% \end{figure*}
% \begin{figure*}
\vspace{-0.3cm}
\begin{align}\label{recv Y including ICI and ISI} 
 { Y}_u[n,m] \! = \!
 \sum\limits_{m' = 0}^{M - 1}  \sum_{c=1}^{N_{R}}
{ H_{n,m}^{u,c}[n \! - \! 1,m']B_c[{n \!- \! 1},m']\tilde X_c[n \!- \!1,m']} 
 %  \notag \\
 %&
+ \! \sum\limits_{m' = 0}^{M - 1}  \sum_{c=1}^{N_{R}} { H_{n,m}^{u,c}[{n},m']B_c[{n },m'] \tilde X_c[{n },m']} 
%  \notag \\
%   \notag \\
% &
+{ W_u}[n,m],
\end{align}
\end{figure*}
where 
$A_{g_{rx},g_{tx}}(\tau ,\nu)=\int_{0}^{T}g_{tx}(t)g^*_{rx} (t-\tau )e^{-j2\pi \nu t}dt$ is the cross-ambiguity function
and $\tilde h_{u,c}(\tau,\nu)$ is the equivalent DD-domain channel response from  $c$-th RF chain of BS to the $u$-th user. 
$\tilde h_{u,c}(\tau,\nu)$ can be derived as \eqref{h_tau_v2} at the top of the next page
by integrating \eqref{ttd ps resp} and \eqref{h_tau_v} with $a=(d-1)N_P +s$.

Without loss of generality, both $g_{tx}(t)$ and $g_{rx}(t)$ are assumed to be rectangular pulses.
Then, as described in \cite{OTFS_r_1}, the cross-ambiguity function in \eqref{Hp}
%$A_{g_{rx},g_{tx}}((n-n')T-\tau ,(m-m')\Delta f-\nu)$
 can be considered non-zero only for $n'=n$ or $n'=[n-1]_N$. 
 Therefore, ${ Y}_u[n,m]$ can be reformulated as \eqref{recv Y including ICI and ISI},
where 
the first item represents the ISI,  the second item denotes ICI, and the third item corresponds to AWGN. By using SFFT and combining  \eqref{DD comp}, \eqref{Xp tf domin}, \eqref{TF comp}, \eqref{Hp} and \eqref{h_tau_v2}, 
the input-output relationship in DD domain is written as
\begin{align}\label{IO_relation}
y_{\! u} \! [k,\ell] \! \! = \! \! \frac{1}{NM} \! \!  \sum_{k'=0}^{N-1}  \!  \sum_{\ell'=0}^{M-1} \!  \sum_{u'=1}^{U} \!  h_{k\! ,\ell}^{u\! ,u'}\! [k'\! ,\ell'] x_{u'}[k'\! ,\ell']\! \!  +\!  \! w_u[k\! ,\ell],
\end{align}
where $h_{k,\ell}^{u,u'}[k',\ell']$ is given by \eqref{H_simple} with $a=(d-1)N_P+s$. 
\begin{figure*}[!htbp]
\vspace{-0.8cm}
{
    \begin{align}\label{H_simple}
 &    h_{k,\ell}^{u,u'}[k',\ell'] 
\! = \! {\frac{{\sqrt{\rho _{c}}}}{{MN}}}  \!
\sum\limits_{n = 0}^{N-1} 
\sum\limits_{m = 0}^{M-1}
  {\sum\limits_{n' = 0}^{N - 1}  \sum\limits_{d = 1}^{N_T} \sum\limits_{s = 1}^{N_P}
\sum_{c = 1}^{N_R} 
{
 \!\int_{\nu}^{} 
 \!\int_{\tau}^{} \!
 \sum\limits_{p = 1}^{P} 
 {{{\alpha }_{p,u}} \! \left| {{\mu _{p,u}}} \right|
 {e^{j2\pi \left( \!- \!\frac{{nk}}{N} \! + \! \frac{{ml}}{M}\right)}}
{e^{j2\pi {\mu _{p,u}}( \! \tau \! - \! t_{c,d} \! - \! \tau _{p,u} \! \! - \! ( \! a \! - \! 1)\frac{{{\psi _{p,u}}}}{{{f_c}}})(v \! - \! {v_{p,u}})}} }} }
{e^{ - j2\pi {\Psi _{c ,d ,s}}}}
\notag \\ 
&   \times \!
 e^{-j2\pi f_c \left({t_{c ,d}}+(a \! - \! 1)\frac{\psi _{p,s}}{f_c} \right)}
\frac{1}{M}\sum\limits_{i\in Q} {{e^{j\frac{{2\pi i(\nu T - m)}}{M}}}} \sum\limits_{m' = 0}^{M - 1} {{e^{j2\pi \frac{{m'(i- \ell'  )}}{M}}}} {e^{j2\pi \frac{{n'k'}}{N}}} B_c[n',m'] D_{c,u'}[k',\ell']
{e^{j2\pi (\nu - m\Delta f)(\tau  + n'T)}}  d\tau dv
    \end{align}
    }
\end{figure*}
The derivation is similar to
 \cite{squint_3,OTFS_r_1}, 
where we apply
\begin{align}
  {A_{{g_{rx}},{g_{tx}}}}(t,f) \approx& \frac{1}{M}\sum\limits_{i \in Q} {{e^{ - j2\pi f\frac{{i T}}{M}}}} ,
\end{align}
with
\begin{align}
  Q =& \left\{ \begin{array}{l}
\{ i \in \mathbb{N} :0 \le \frac{{i T}}{M} < t + T\} ,t \in ( - T,0)\\
\{ i \in \mathbb{N} :t \le \frac{{i T}}{M} < T\} ,t \in [0,T)
\end{array} \right. . 
\end{align}

\begin{figure*}[!htbp]
\vspace{-0.8cm}
\begin{align}\label{H_p_l from dd domain to tf domain}
\tilde{h}_{k,\ell}^{u,u^{\prime}}\left[k^{\prime},\ell^{\prime}\right]  = & \sum_{n=0}^{N-1} \sum_{m=0}^{M-1} \sum_{p=1}^{P} \sum_{c=1}^{N_{R}} \sum_{d=1}^{N_{T}} \sum_{s=1}^{N_{P}} \frac{\sqrt{\rho_{c}}}{NM} \alpha_{p,u} e^{j2\pi\left(\frac{n(k^{\prime}-k)}{N} - \frac{m(\ell^{\prime}-\ell)}{M}\right)} e^{-j2\pi\left(\left((d-1)N_{P}+s-1\right)\psi_{p,u} + t_{c,d}f_{c}\right)\left(1 + \frac{m\Delta f}{f_{c}}\right)} e^{j2\pi\Psi_{c,d,s}}  \notag \\
& \times 
% e^{-j2\pi f_c \left({t_{c ,d}}+(a-1)\frac{\psi _{p,s}}{f_c} \right)} 
e^{j2\pi\nu_{p,u}\frac{\left(\ell^{\prime} + \tilde{\ell}_{c,d} + \ell_{p,u} + \left((d-1)N_{P}+s-1\right)\frac{\psi_{p,u}}{f_{c}T_{s}}\right)T}{M}}
% \notag \\
% & 
\times 
\left\{ \begin{array}{l} e^{-j2\pi\frac{m\ell_{p,u}}{M}} e^{j2\pi n\left(\frac{k_{p,u}}{N} + \frac{m}{\mu_{p,u}}\right)}, \ell^{\prime} \in \mathcal{L}_{ICI}^{p,u}  \\ e^{-j2\pi\frac{k^{\prime}}{N}} e^{-j2\pi\frac{m\ell_{p,u}}{M}} e^{j2\pi\left(n-1\right)\left(\frac{k_{p,u}}{N} + \frac{m}{\mu_{p,u}}\right)}, \ell^{\prime} \in \mathcal{L}_{ISI}^{p,u} \end{array} \right.,
\end{align}
\hrulefill
\end{figure*}

%the $ h^{u,u'}_{k,\ell}[k',\ell']$ 
%related to ${{ H}_{n,m}^{u}[n,m']}$ and ${{ H}_{n,m}^{u}[n-1.m']}$, respectively. 
%The proof has been omitted due to its similarity with the derivation in Appendix \ref{proof of bd after precoding}, which corresponds to Section IV.

\subsection{Discussion of the Doubly Squint Effect }

In order to illustrate the doubly squint effect in wide-band massive MIMO-OTFS scenarios, we first investigate the equivalent delay-Doppler-domain channel response without involving the digital precoding (e.g., \( B_{c}[n,m]=1 \) and \( D_{c,u}[k,l]=1 \), \( \forall c, u, m, n, k, l \)). Thus, the DD-domain equivalent channel response can be simplified to \( \tilde{h}_{k,\ell}^{u,u^{\prime}}[k^{\prime},\ell^{\prime}] \) as shown in (20), where
$k_{p,u} = \nu_{p,u}NT$,  { $\ell_{p,u}=\frac{\tau_{p,u}M}{T}$}
\( \tilde{\ell}_{c, d} = \frac{t_{c, d}}{T_{s}}, \, \mu_{p, u} = \frac{f_{c}}{\nu_{p, u}} \) and \( T_{s} \) is the sampling period. \( \mathcal{L}_{ICI}^{p,u} \) and \( \mathcal{L}_{ISI}^{p,u} \) are the index sets of the \( p \)-th path with \( u \)-th user for ICI and ISI, respectively.

\emph{Proof:} Appendix \ref{proof of huu}.

\begin{figure}[htbp]
  \centering
  %\label{DDA channel}
  % Requires \usepackage{graphicx} and \usepackage{subfigure}
  \subfigure[]
  {\label{new_patterns2}
    \begin{minipage}{0.22\textwidth} % 设置子图宽度为页面宽度的45%
    \centering
    \includegraphics[width=\linewidth]{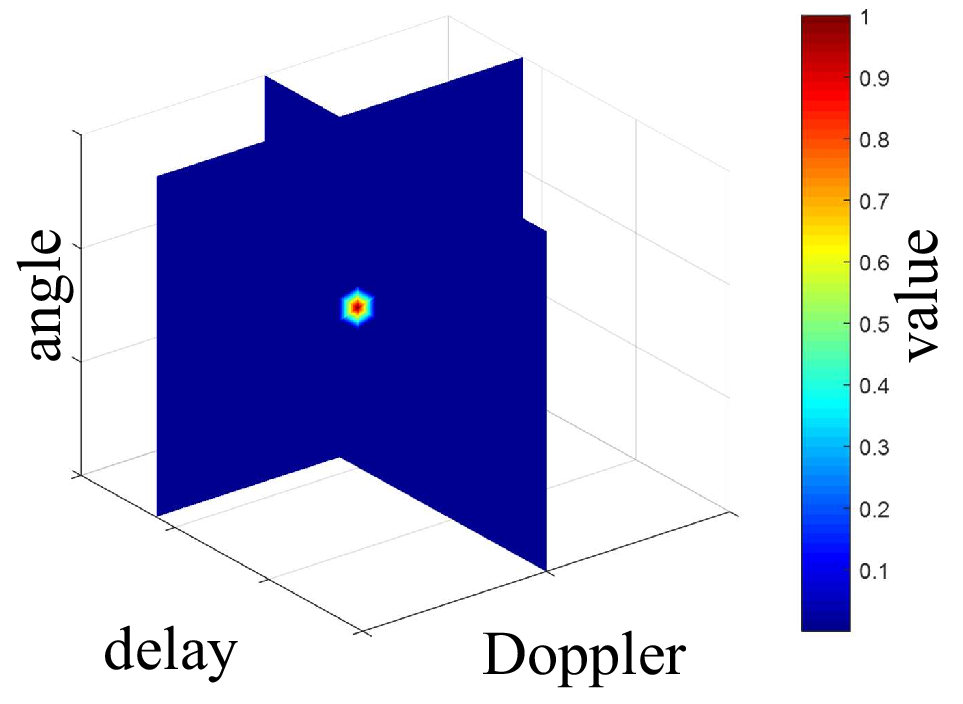} % 子图宽度设置为行宽
    \end{minipage}
  }
  \hfill % 填充空白，分隔两个子图
  \subfigure[]
  {\label{new_patterns1}
    \begin{minipage}{0.22\textwidth} % 设置子图宽度为页面宽度的45%
    \centering
    \includegraphics[width=\linewidth]{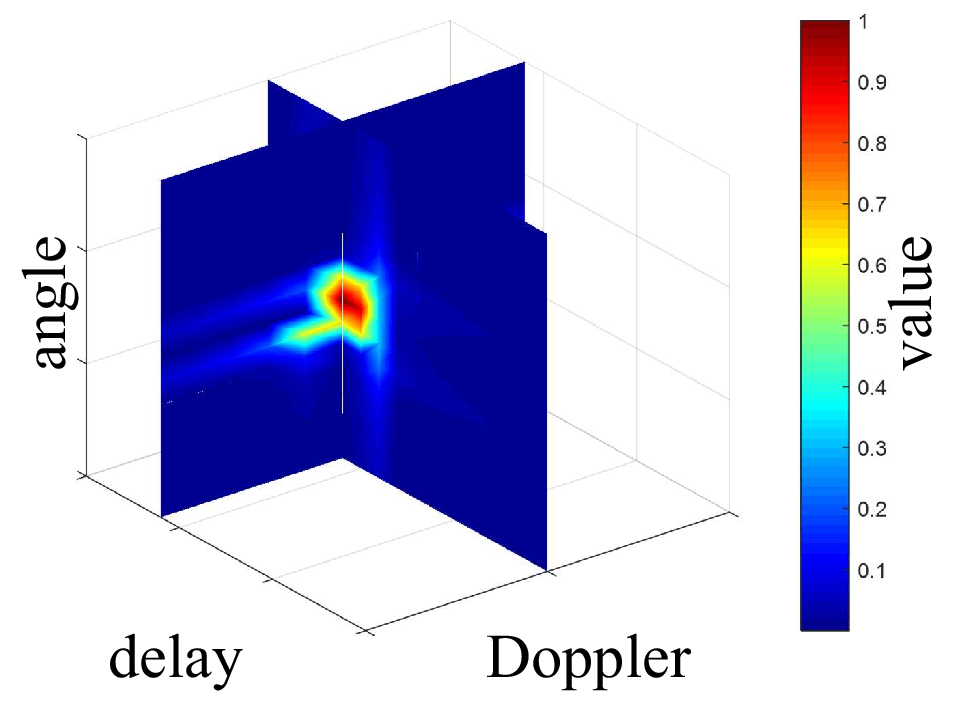} % 子图宽度设置为行宽
    \end{minipage}
  }
  \caption{Delay-Doppler-angle-domain channel  response (a) in narrowband system; (b)  in wideband system.}
   \label{DDA channel}
\end{figure}

Equation  \eqref{H_p_l from dd domain to tf domain}  reveals that the DD-domain equivalent channel response consists of space-frequency-angle coupling $e^{\frac{-j2\pi((d-1)N_{P}+s-1)\psi_{p,u}m\Delta f}{f_{c}}}$ and time-frequency-Doppler coupling $e^{\frac{j2\pi mn}{\mu_{p,u}}}$, leading to beam squint effect and Doppler squint effect, respectively. 
{
The coupling $e^{\frac{-j2\pi((d-1)N_{P}+s-1)\psi_{p,u}m\Delta f}{f_{c}}}$ indicates that the beam squint effect introduces a phase shift depends on subcarrier and antenna.
Furthermore, this coupling indicates that the beam squint effect is influenced by the number of antennas $N_A$, the spatial angle $\psi_{p,u}$ and the ratio of bandwidth to carrier frequency $\frac{B}{f_c}$. 
Thus, the beam squint effect increases when either
magnitude of the spatial angle $|\psi_{p,u}|$ or the ratio $\frac{B}{f_c}$ increases.
The time-frequency-Doppler coupling $e^{\frac{j 2 \pi m n}{\mu_{p,u}}}$ indicates that the Doppler squint effect
introduces a subcarrier-dependent phase shift. 
Considering $\mu_{p,u} = \frac{f_c}{\nu_{p,u}} = \frac{v_l}{v_{p,u}}$, the maximum Doppler squint effect can be expressed as $\frac{M N \nu_{p,u}}{f_c} = \frac{M N v_{p,u}}{v_l}$. Therefore, the Doppler squint effect depends on the radial velocity $v_{p,u}$ and the product $M \times N$.
In practical systems, the duration of an OTFS frame is given by $T_{\text{symbol}} = N T = \frac{N M}{B}$. Consequently, the maximum Doppler squint effect can be represented as $e^{j 2 \pi \frac{v_{p,u} B T_{\text{symbol}}}{v_l}}$.
Thus, the Doppler squint effect is determined by the radial velocity, bandwitdh and symbol duration.
Therefore, in wideband massive MIMO-OTFS systems with high mobility, the combined impact of Doppler squint and beam squint effect, referred to as the doubly squint effect, must be considered.
% Given $B = M \Delta f$, the beam squint effect is influenced by three key factors: the number of antennas $N_A$, the spatial angle $\psi_{p,u}$, and the ratio of bandwidth to carrier frequency $\frac{B}{f_c}$. Specifically, the beam squint effect increases with the magnitude of the spatial angle $|\psi_{p,u}|$ and the ratio $\frac{B}{f_c}$.
% The time-frequency-Doppler coupling term $e^{\frac{j 2 \pi m n}{\mu_{p,u}}}$ introduces a subcarrier-dependent phase shift due to the Doppler squint effect, which becomes more pronounced as the parameter $n$ increases. Given $\mu_{p,u} = \frac{f_c}{\nu_{p,u}} = \frac{v_l}{v_{p,u}}$, the maximum Doppler squint effect can be expressed as $\frac{M N \nu_{p,u}}{f_c} = \frac{M N v_{p,u}}{v_l}$. Therefore, the Doppler squint effect depends on the radial velocity $v_{p,u}$ and the product of the number of subcarriers $M$ and the number of OFDM symbols $N$.
% Additionally, the duration of an OTFS frame is given by $T_{\text{symbol}} = N T = \frac{N M}{B}$. Consequently, the maximum Doppler squint effect can be represented as $e^{j 2 \pi \frac{v_{p,u} B T_{\text{symbol}}}{v_l}}$.
% In communication systems, symbol time is typically constrained to prevent channel aging. From the above analysis, we conclude that both Doppler squint and beam squint are influenced by the bandwidth. Therefore, in wideband communication systems, the combined effect of Doppler squint and beam squint, referred to as the doubly squint effect, must be considered.

In massive MIMO-OTFS systems, we can characterize the channel in the delay-Doppler-angle domain.}
By using discrete Fourier transform (DFT) on $\tilde{h}_{k,\ell}^{u, u^{\prime}}\left[k^{\prime},\ell^{\prime}\right]$ with respect to the antenna index \(a\), we can derive the delay-Doppler-angle-domain channel response as $\tilde{h}_{k,\ell,\psi}^{u, u^{\prime}}\left[k^{\prime},\ell^{\prime}\right] = \frac{1}{\sqrt{N_A}}\sum_{a=1}^{N_A}\tilde{h}_{k,\ell}^{u,u'}[k',\ell'] e^{j2\pi\frac{a\psi}{N_A}}$, where \(a = (d-1)N_P + s\) and
$\psi$
is the spatial angle in angle domain. 
As illustrated in \figurename{ \ref{DDA channel}},
in narrowband and small array systems, i.e., $\frac{N_{A} m\Delta f}{f_{c}} \ll 1$ and $\frac{m n}{\mu_{p, u}} \ll 1$, the impact of space-frequency-angle coupling and time-frequency-Doppler coupling terms can be disregarded.
However, in wideband massive MIMO-OTFS systems, these two coupling terms can lead to dispersion within the delay-Doppler-angle domain.
% as illustrated in \figurename{ \ref{DDA channel}}. 
Such dispersion can subsequently cause additional inter-path interference, ICI and ISI. 
In this work, we propose to compensate for these coupling effects by optimizing digital precoding (i.e., $D_{c,u}[k,\ell]$ and $B_c[n,m]$) and analog precoding
(i.e., $t_{c,d}$ and $\Psi_{c,d,s}$) in Section IV.

\section{Proposed Uplink Channel Estimation With Chirp-based Pilot}

\begin{figure}[htbp]
 \centering
 \includegraphics[width=75mm]{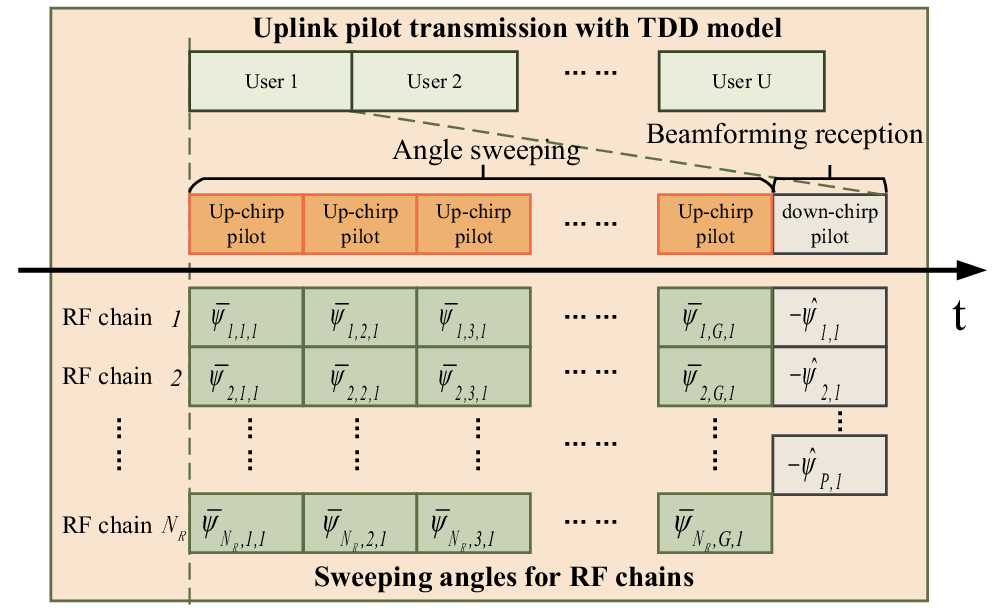}
 \caption{Proposed channel estimation scheme with TDD model.}
 \label{transmission_model}
\end{figure}

Before optimizing digital percoding and analog precoding, it is essential and necessary to obtain the channel parameters accurately. 
{
In OTFS channel estimation, embedded pilot aided estimation and preamble aided estimation are two feasible schemes.   
The embedded pilot scheme integrates pilot symbols into the DD domain grid directly.  
The receiver extracts the pilot to estimate the channel state information, which is then used for data detection \cite{emb_pilot}. 
% However,  the difference in power allocation between pilots and data can lead to a high peak-to-average power ratio  (PAPR), which places high demands on the power amplifier (PA) of the transmitter.
The preamble-based estimation scheme inserts a known pilot sequence (i.e., preamble) at the beginning of data frame. 
The receiver uses this preamble to estimate the channel state, which can be used for date detection or precoding.
In this work, channel estimation is not for data detection but rather for precoding design. 
Thus, preamble-based channel estimation is particularly suitable for this scenario.

In traditional multiuser MIMO-OTFS channel estimation, the pilots of different users are usually assigned in same DD domain symbol to realize simultaneous multiuser channel estimation \cite{MUMIMO_CE}. 
However, it is infeasible in massive MIMO system with hybrid structure as the base station cannot receive the pilot in the DD domain if the angle of the user is unknown. 
In this paper, we propose an chirp-based channel estimation method in time division duplex (TDD)  mode with angle sweeping.
}
The chirp signal, which has superior cross-correlation properties, is widely used in radar systems.
In addition, the chirp signal exhibits excellent pulse compression characteristics, thereby leading to superior estimation performance of range and Doppler \cite{ISAC_0,LFM_1}.
{
Leveraging these advantages, 
the chirp-based pilot is selected as the preamble for our channel estimation.
Specifically,
we first preset an angle-domain codebook with $N_A$ angles. 
Each RF chain can independently receive the signal from a specific angle by controlling the PSs and TTDs.
Within a TDD slot, the BS can receive signals from $N_R$ angles simultaneously.
\figurename{ \ref{transmission_model}}  illustrates the structure of the proposed channel estimation scheme, which is comprised
of two phases.
The first phase is angle sweeping, during which each user sends $G$ up-chirp pilots across $G = \left\lceil \frac{N_A}{N_R} \right\rceil $ separate TDD slots.
}
The BS adjusts the sweeping angle of each RF chain.
Following angle sweeping, the BS can derive the angle of the paths with the angle of the $p$-th path denoted as $\hat \psi_{p,u}$.
The second phase is beamforming receiving, where the user transmits a down-chirp pilot in $(G+1)$-th slot.
The BS RF chains align the sweeping angles to the angles estimated in angle sweeping phase. 
Then, the BS can receive down-chirp pilot from $P$ scattering paths directly within a TDD slot.
Consequently, the BS can calculate the DFT-angle sequences from up-chirp and down-chirp pilots. 
Finally,the channel parameters can be derived with DFT-angle sequences.

In this work, the up-chirp pilot $\acute C[i]$ and down-chirp pilot $\grave C[i]$ are respectively represented as
\begin{align}\label{upchirp samp} 
\acute C[i]
%= \acute {\rm C}(t)|_{t=iT_s}  
%\notag \\ 
= e^{j2\pi \kappa i^2}, i=0,\dots,M-1,  \\
\grave C[i]
%= \grave {\rm C}(t)|_{t=iT_s}  
%\notag \\ 
= e^{- j2\pi \kappa i^2}, i=0,\dots,M-1, 
\end{align} 
where  $M$ is the length of chirp sequence and $\kappa = \frac{1}{2M}$ is chirp rate \cite{chirp1, chirp2},.
To mitigate the effects of the multipath, 
 a  chirp-periodic prefix (CPP) is introduced \cite{AFDM_1} as
\begin{align}\label{cpp}
&\acute C[i]\! =\!  \acute C[\!  M\! 
 +\!  i] e^{j2\pi \kappa(M^2\! 
 + \! 2Mi)}\! ,\!  
 %\notag \\
 i\!  =\!  -\! N_{CPP},\ldots,\! -\! 2,\! -\! 1, \\
&\grave C[i] \! =\!  \grave C[\! M \! + \! i] e^{-j2\pi \kappa(M^2 \! 
 + \! 2Mi)}\!  , 
 %\notag \\
 i\!  =\! \!  -\!  N_{CPP},\ldots,\! -\! 2,\! -\! 1,
\end{align}
%for  $n=-N_{ccp},\ldots,-1$,
 where $N_{CPP}$ is the length of the CPP and surpasses the maximum  delay spread. 
In the subsequent, we will introduce angle sweeping and beamforming receiving phases, respectively.

\subsection{Angle Sweeping for Up-chirp Pilot}

\begin{figure*}[!htbp]
 \centering
 \includegraphics[width=170mm]{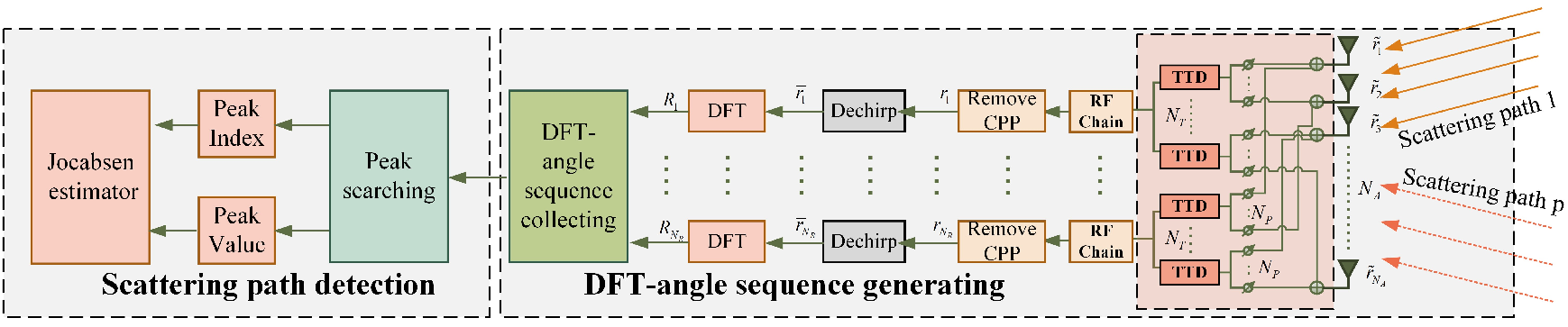}
 \caption{Angle sweeping process for up-chirp pilot.}
 \label{Channel estimation structure}
\end{figure*}

In the first phase, each user transmits an up-chirp pilot at each TDD slot and BS receives pilot from $P$ scattering paths. 
Taking the $g$-th TDD slot with $u$-th user as an example, the up-chirp pilot received by the $a$-th antenna of the BS can be represented as
\begin{align}\label{p-th chirp in r antenna}
&\tilde {r}_{a,g,u}[i]=  \sum_{p=1}^{P} { \alpha^g_{p,u} }e^{j2\pi \kappa\left(i- \left(\ell^g_{p,u} + (a-1)\frac{\psi_{p,u}}{f_cT_s } -\frac{\nu_{p,u}}{f_c}i\right)\right)^2 }
 \notag \\
& \times \!
e^{\! \! -j2\pi \left( \!
 (a\! -\! 1)\frac{\psi_{p,u}}{f_c} \! -\!  \nu_{p,u}i T_s \!\right)}\! \! + \! \tilde w_{a,g,u}[i], 
  % \notag \\
 i \! = \! 0, \!1,\dots,M\! \!- \!1,
\end{align}
%The BS receives signals with different angle of arrival (AoA) by independently tuning the combiner according to the scanning angle set
%$\mathbf{S} = \{\psi_{n_{RF}}  | \psi_{n_{RF}} = -\frac{1}{2} + \frac{\varphi_{n_{RF}}}{R}, \varphi_{n_{RF}} = 0, 1, \ldots, R - 1\}$, where $\psi_{n_{RF}} $ corresponds to the normalized scanning angles.+
where  $\alpha^g_{p,u}  = \tilde \alpha_{p,u} e^{-j2\pi f_c \left (\tau_{p,u} - \frac{\nu_{p,u}}{f_c } (g-1)(M+N_{CCP})T_s \right)  }  $ is the path gain at the $g$-th slot and  $\tilde w_{a,g,u}[i]$ is the AWGN at the $a$-th BS antenna for the estimation period of the $u$-th user. 
\footnote{
% In \cite{squint_3}, Wang \emph{et al.} illustrated that
Doppler squint effect can be represented as a time-varying delay in the time-domain channel response \cite{squint_3}, (i.e., $ h( t) = \sum_{p=1}^{P} \alpha_p(t) \delta(t - \tau_p(t))$, where $\tau_p(t) = \tau_p - \frac{\nu_p}{f_c}t$). 
Taking into account the effect of time-varying delay, the gain and delay in each TDD slot should be considered different.
}
 $\ell^g_{p,u}$ is the delay  at the $g$-th slot and can be expressed as
\begin{align} \label{delay_at_g}
\ell^g_{p,u} = \ell_{p,u} - \frac{\nu_{p,u}}{f_c} (M+N_{CCP}) (g-1).
\end{align}
% \emph{Proof:} Appendix \ref{time-varying delay}.
The processes illustrated in \figurename{ \ref{Channel estimation structure}} is then implemented.

\subsubsection{DFT-angle Sequence Generating}
With respect to the $c$-th RF chain, the BS adjusts the sweeping angle of this RF chain by optimizing the TTD set $\mathbf{T}_{c}$ and PS set $\mathbf{P}_{c} \triangleq \cup_{d=1}^{N_{T}} \mathbf{P}_{c, d}$. 
Specifically, during the $g$-th slot with $u$-th user, the sweeping angle of the $c$-th RF chain is represented as $\bar{\psi}_{c, g, u} \in \mathbf{A}$.
% with $\bar{\psi}_{c, g, u} \neq \bar{\psi}_{c^{\prime}, g^{\prime},u}$ either when $c \neq c^{\prime}$ or $g \neq g^{\prime}$. 
The angle set $\mathbf{A} $ is defined as $\mathbf{A} = \left\{\bar{\psi} \mid \bar{\psi} = -\frac{1}{2} + \frac{\bar{\varphi}}{N_S}, \bar{\varphi} = 0, 1, \ldots, N_S-1 \right\}$ with $N_S = GN_R$. 
The BS can subsequently sweep the signal from the sweeping angle.

We first optimize the TTD and PS 
to improve the sweeping performance for DFT-angle sequence.
% With respected to the $c$-th RF chain at the $g$-th TDD slot,
After the processing of PS set, TTD set, and CPP removal, the output signal  with sweeping angle $\bar \psi_{c,g,u}$
is given by
\begin{align}
&r_{c,g,u}[i,\bar \psi_{c, g,u}] = \sum_{p=1}^{P} \sum_{d=1}^{N_{T}} \sum_{s=1}^{N_{P}} {  \alpha^g_{p,u} } e^{-j 2\pi f_{c}\left((a-1)\frac{\psi_{p, u}}{f c} + t^g_{c, d}\right)} 
\notag \\
&\times e^{j 2\pi \kappa\left(i - \left(\frac{t_{c, d}^{g}}{T_{s}} + \ell^g_{p, u} + \left((d-1) N_{P} + s - 1\right) \frac{\psi_{p, u}}{f_{c} T_{s}} - \frac{\nu_{p, u}}{f_{c}} i\right)\right)^{2}}
\notag \\ 
&\times e^{j 2\pi i \nu_{p, u} T_{s}}  e^{j 2\pi \Psi_{c, d, s}^{g}} + w_{c, g,u}[i], i = 0, \ldots, M-1, 
\end{align} 
where $w_{c,g,u}[i] = \sum\limits_{d  = 1}^{N_T} 
\sum\limits_{ s  = 1}^{N_P} \tilde w_{a,g,u}[i]* \bar {\rm h}_{c,d,s}[i]$ and $\bar{h}_{c, d, s}[i] =  \bar{h}_{c, d, s}[t] \left|_{t = i T_{s}, \Psi_{c, d, s} = \Psi_{c, d, s}^{g}, t_{c, d} = t_{c, d}^{g}} \right.$ with $a \!= \!(d \!- \!1)N_P+s$. 
% We omit the subscripts $c$ and $g$ of $r_u[\cdot]$ for simplicity.
The parameters $\Psi_{c,d,s}^g$ and $t_{c,d}^g$ respectively represent the values of $\Psi_{c,d,s}$ and the $t_{c,d}$ at the $g$-th slot.
The de-chirp sequence is then derived by
\begin{align}\label{p-th de-chirp in r antenna up} 
&\bar{r}_{{c,g,u}}[i,\bar \psi_{c, g,u}] =  r_{{c,g,u}}[i,\bar \psi_{c, g,u}] \times e^{-j 2\pi \kappa i^2}
\notag \\
&\approx  \sum_{p=1}^{P} \sum_{d=1}^{N_{T}} \sum_{s=1}^{N_{P}} {  \alpha^g_{p,u} } e^{j 2\pi \kappa \left( \frac{t_{c, d}^{g}}{T_{s}} + \ell^g_{p, u} + \left( (d-1) N_{P} + s - 1 \right) \frac{\psi_{p, u}}{f_{c} T_{s}} \right)^2} \notag \\
& \times e^{j 2\pi i \left( \nu_{p, u} T_{s} - 2\kappa \left( \frac{t_{c, d}^{g}}{T_{s}} + \ell^g_{p, u} + \left( (d-1) N_{P} + s - 1 \right) \frac{\psi_{p, u}}{f_{c} T_{s}} \right) \right)} \notag \\
& \times e^{-j 2\pi \left( \left( (d-1) N_{P} + s - 1 \right) \psi_{p, u} + f_{c} t_{c, d}^{g} - \Psi_{c, d, s}^{g} \right)} + \bar{w}_{c, g,u}[i],
\end{align}
where  $\bar w_{c,g,u}[i] = w_{c,g,u}[i]\times e^{-j2\pi \kappa i ^2} $.
The use of the approximation is based on the condition that 
 $\kappa i \nu_{p,u}/f_c$ is typically negligible, $ \forall i=0,1,\dots,M-1$. 
 {
By $N_A$  sweepings and using the DFT  on $\bar r_{c,g,u}[i,\bar{\psi}_{c,g,u}]$ with time $i$, we can derive the DFT-angle sequence  $R'_{c,g,u}[m,{\bar \psi_{c,g,u}}]$as \eqref{p-th dft in r antenna appendix} at the top of the next page, where $R'_{c,g,u}[m,{\bar \psi_{c,g,u}}]$ is divided by $M$ to normalize, $T_c = 1/f_c$,  $m=-M/2,-M/2+1,\dots,M/2-1$ and $W_{c,g,u}[m]$ is the DFT of $\bar w_{c,g,u}[i]$ with time $i$.
} 
\begin{figure*}[!htbp]
{
\begin{align}\label{p-th dft in r antenna appendix}
 & R' _{c,g,u}[m,\bar \psi_{c, g,u}]
\approx  
%\frac{1}{N} 
\sum\limits_{p= 1}^{P} \sum\limits_{d= 1}^{N_T} 
\sum\limits_{s = 1}^{N_P} { \alpha^g_{p,u} }
e^{-j2\pi (f_c t^g_{c,d}+(d-1)N_P\psi_{p,u})\left(1 - 
\left((f_c t^g_{c,d} + (d-1)N_P\psi_{p,u})\frac{1}{f_cT_s} + 2{\ell^g_{p,u}}  \right) \frac{\kappa}{f_cT_s}  \right)}
\notag \\
% &\times 
%  \notag \\
&\times
e^{j2\pi\kappa (\ell_{p,u}^g)^2}
\frac{\sin \left(-\pi \left(m-M\nu_{p,u} T_s + 2M\kappa{\left({\ell^g_{p,u}} + ({f_c t^g_{c,d}} +(d-1)N_P{\psi_{p,u}})\frac{1}{f_c T_s} + {(s-1)\frac{\psi _{p,u}}{f_c T_s}}\right)}\right) \right)}
{M\sin\left(-\frac{\pi}{M}\left(m-M\nu_{p,u} T_s + 2M\kappa{\left({\ell^g_{p,u}} + ({f_c t^g_{c,d}} +(d-1)N_P{\psi_{p,u}})\frac{1}{f_cT_s} + {(s-1)\frac{\psi _{p,u}}{f_cT_s}}\right)}\right)\right)}
\notag \\
&\times
 e^{-j\pi \frac{M-1}{M} \left(m-M\nu_{p,u} T_s + 2M\kappa{\left({\ell^g_{p,u}} + ({f_c t^g_{c,d}} +(d-1)N_P{\psi_{p,u}})\frac{1}{f_cT_s}  \right)}  \right)}
 e^{-j2\pi \left(-\Psi^g_{c,d,s} + (s-1) \left({\psi_{p,u}} + (M-1)\frac{\kappa \psi_{p,u}}{f_c T_s} \right) \right)}
+ W_{c,g,u}[m],
\end{align}
}
% \hrulefill
\end{figure*}

{
\subsubsection{ TDD-PS Net Design  and Scattering Path Detection}
The BS determines the existence of a detection path by comparing the energy from the sweeping angle to a predefined threshold $\eta$.
Without loss of generality, the angle of each scattering path is assumed to be unique for a specific user.
As illustrated in , the BS can derive the detected angle set $\mathbf{Q}_u$ if the
the normalized peak $R'_{c,g,u}[m,\bar \psi_{c,g,u}]$
exceeds the threshold $\eta$ as
\begin{align}\label{energy dedection}
%X = \max \{ x_i : x_i \in S, x_i \geq \theta \} 
\mathbf{Q}_u  = \left \{\bar \psi_{c,g,u}
 \left |
 \frac{\max \left( \left|R'_{c,g,u}[m,{\bar \psi_{c,g,u}}] \right|
 \right)}{ { \zeta}_u[\bar \psi_{c,g,u}] }
 > \eta  
  \right.
  ,\bar \psi_{c,g,u} \in \mathbf{A}
 \right \},
\end{align}
where $\eta$ is related to the probability of false alarm and miss rate \cite{alarm_prob}.
$\zeta_u \left[\bar{\psi}_{c, g,u}\right]$ is the energy of received pilot expressed as
\begin{align}\label{energy}
{\zeta_u}[\bar \psi_{c,g,u}] =& \sum_{m=0}^{M-1} \left|R'_{c,g,u}[m,{\bar \psi_{c,g,u}}]\right|^2.
\end{align}
After $G$ TDD slots, $\mathbf{Q}_u$ can be obtained by sweeping all angles in the set $\mathbf{A}$. We denote $P'$ as the number of elements in set $\mathbf{Q}_u $. Then, $P'$ scattering paths can be detected after angle sweeping.
However, the configuration of TTD and PS are still unknown for our receiving processing.
Thus, We first analyze the configuration of  TTD and PS such that the signal energy is maximized when the received signal comes from ${\psi_{p,u}} = -\bar \psi_{c,g,u}$.
Specifically, from the $p$-th scattering path in \eqref{p-th dft in r antenna appendix}, 
the value $\max \left( \left|R'_{c,g,u}[m,{\bar \psi_{c,g,u}}] \right|
 \right)$
 can be maximized if TTD delay $t_{c,d}^g$ and PS phase shift $\Psi_{c,d,s}^g$ is given by
% \begin{align}\label{ttd in est ch}
% {t^g_{c ,d}} =& - (d-1) {N _P}{\psi _{p',u}}/{f_c},
% \\
% \label{ps in est ch}
%  {\Psi^g_{c ,d ,s}} =&
%  (s-1) \left ( 1 
%  + \frac{\kappa  (M-1)}{T_sf_c}\right){\psi _{p',u}}.
%   \end{align}
% Consequently,  \eqref{ttd in est ch} and \eqref{ps in est ch} can be extended to  universal forms for scanning angles ${\bar \psi _{c,g,u}}\in \mathbf{A}$ as 
\begin{align}\label{ttd in est set}
{t^g_{c ,d}} =& \tilde{t} + (d-1) {N _P}{\bar \psi _{c,g,u}}/{f_c}, \\
\label{ps in est set}
 {\Psi^g_{c ,d ,s}} =& 
 (1-s)  \left ( 1 
 + \frac{\kappa  (M-1)}{T_sf_c} \right){\bar \psi _{c,g,u}},
  \end{align}
where ${\tilde t}$ is a constant to ensure $t_{c,d}$ is positive.
% Subsequently, we can derive the set $\mathbf{Q}_u $ after $N_A$ sweeping angles are detected during $G$ TDD slots. 
From \eqref{p-th dft in r antenna appendix},
there is a summation over the Dirichlet Sinc function about $d$ and $s$ in $R'_{c ,g,u}[m, \bar \psi_{c,g, u}]$.
Considering the power-focusing property of the Sinc function, a peak can be obtained only when the deviation between sweeping angle and real angle of scattering path angle is sufficiently small.
Given detected angle $\bar \psi_{c,g, u}\in \mathbf{Q}_u$, the BS can derive the peak index by searching $\left|R[m,\bar \psi_{c,g, u}]\right|$ for $m=-M/2,-M/2-1,\dots,M/2-1$, i.e.,
 \begin{align}
m_{c,g,u}^{\nearrow } = & 
 \! \! \! \! \! \!  \! \! \! \! \! \!\!\mathop{\arg\max}\limits_{m=-M/2,-M/2+1,\dots,M/2-1} 
\! \! \! \! \! \! \! \! \! \! \! \! \!
\left | {{R}}[m, \bar \psi_{c,g, u}]
\right |,
\bar \psi_{c,g, u}\in \mathbf{Q}_u
%\notag \\
%\quad \bar \psi_{a,g}\in \mathbf{Q}.
 \end{align}
In order to enhance the performance of the estimation, we employ the Jocabsen estimator \cite{Jocabsen} 
to derive more refined  angle  and peak index  as shown in \eqref{est angle s} and \eqref{est m}, expressed at the top of this page.
\begin{figure*}[!htbp]
\vspace{-0.8cm}
{
\begin{align}\label{est angle s}
 J ({ m_{c,g,u}^\nearrow },{\bar \psi _{c,g,u}}) \! = \! 
 {\Re} \! \left \{ \! {\frac{{ {{R_{c,g,u}}}[m_{c,g,u}^\nearrow ,\bar \psi_{c,g,u}+\frac{1}{N_S}]- {{R_{c,g,u}}}[m_{c,g,u}^\nearrow ,\bar \psi_{c,g,u}-\frac{1}{N_S}]}}{{2{{R_{c,g,u}}} \![m_{c,g,u}^\nearrow  , \!\bar \psi_{c,g,u}] \! -\! {{R_{c,g,u}}}[m_{c,g,u}^\nearrow ,\! \bar \psi_{c,g,u} \! + \! \frac{1}{N_S}\! ]\!  - \! {{R_{c,g,u}}}[m_{c,g,u}^\nearrow ,\! \bar \psi_{c,g,u}\! -\! \frac{1}{N_S}]}}} \right \} \! \frac{1}{N_S}\! -\!\bar \psi_{c,g,u} ,\! 
% \bar \psi_{c,g,u}  \in \mathbf{Q},
\end{align} 
\begin{align}\label{est m}
J'({ m_{c,g,u}^\nearrow },{\bar \psi _{c,g,u}}) = 
 m_{c,g,u}^\nearrow  - {\Re} \left \{ {\frac{{{{R_{c,g,u}}}[m_{c,g,u}^\nearrow +1,\bar \psi_{c,g,u}]- {{R_{c,g,u}}}[m_{c,g,u}^\nearrow -1,\bar \psi_{c,g,u}]}}{{2{{R_{c,g,u}}}[m_{c,g,u}^\nearrow ,\bar \psi_{c,g,u}] - {{R_{c,g,u}}}[m_{c,g,u}^\nearrow +1,\bar \psi_{c,g,u}] - {{R_{c,g,u}}}[m_{c,g,u}^\nearrow -1,\bar \psi_{c,g,u}]}}} \right \}, 
 % \bar \psi_{c,g,u}  \in \mathbf{Q},
\end{align} 
}
% {\color{blue}
% \begin{align}\label{est angle s}
%  J ({ m_{c,g,u}^\nearrow },{\bar \psi _{c,g,u}}) \! = \! 
%  -\! \left ( \! {\frac{{ \left |{R_{c,g,u}}[m_{c,g,u}^\nearrow ,\bar \psi_{c,g,u}+\frac{1}{N_S}] \right| - \left |{{R_{c,g,u}}}[m_{c,g,u}^\nearrow ,\bar \psi_{c,g,u}-\frac{1}{N_S}] \right|}  }
% {{4  \left| {{R_{c,g,u}}} \![m_{c,g,u}^\nearrow  , \!\bar \psi_{c,g,u}] \right|\! -\! 2 \left| {{R_{c,g,u}}}[m_{c,g,u}^\nearrow ,\! \bar \psi_{c,g,u}  \! + \! \frac{1}{N_S}\! ] \right|\!  - \! 2\left |{{R_{c,g,u}}}[m_{c,g,u}^\nearrow ,\! \bar \psi_{c,g,u}\! -\! \frac{1}{N_S}]\right|} } } \right ) \! \frac{1}{N_S}\! -\!\bar \psi_{c,g,u} ,\! 
% % \bar \psi_{c,g,u}  \in \mathbf{Q},
% \end{align} 
% \begin{align}\label{est m}
% J'({ m_{c,g,u}^\nearrow },{\bar \psi _{c,g,u}}) \! = \!
%  m_{c,g,u}^\nearrow \! + \!\left \{ {\! \!\frac{  { \left |{R_{c,g,u}}[m_{c,g,u}^\nearrow +1,\bar \psi_{c,g,u}] \right |- \left |{R_{c,g,u}}[m_{c,g,u}^\nearrow -1,\bar \psi_{c,g,u}] \right |}}
% {{4  \left |{{R_{c,g,u}}}[m_{c,g,u}^\nearrow ,\bar \psi_{c,g,u}] \right| \!- \!2\left |{{R_{c,g,u}}}[m_{c,g,u}^\nearrow +1,\bar \psi_{c,g,u}] \right| \!- \!2\left |{{R_{c,g,u}}}[m_{c,g,u}^\nearrow \! - \!1,\bar \psi_{c,g,u}] \right|}}} 
% \!  \right \}, 
%  % \bar \psi_{c,g,u}  \in \mathbf{Q},
% \end{align} 
% }
% \hrulefill
\end{figure*}
For the $u$-th user, we denote 
 $\mathbf{\Phi}_u = \{\hat \psi _{c,g,u}\left| \hat \psi _{c,g,u}=J({ m_{c,g,u}^\nearrow },{\bar \psi _{c,g,u}}), \bar \psi_{c,g,u} \in \mathbf{Q}_u\right.\}$
 and
 $\mathbf{m}_u= \{\hat { m}_{c,g,u}^\nearrow  \left | \hat { m}_{c,g,u}^\nearrow = J'({ m}_{c,g,u}^\nearrow ,\bar \psi_{c,g,u}),\bar \psi_{c,g,u} \in \mathbf{Q}_u   \right.\}$
 as angle set and peak index set, respectively.
For the sake of illustration, we denote the actual scattering path angle closest to sweeping angle $-{\bar \psi _{c,g,u}}$ as $\psi_{p',u}$.
Specifically, 
the sets $\mathbf{\Phi}_u$ and $ \mathbf{m}_u^\nearrow$ can be respectively represented as
\begin{align}
 \label{set phi up}
 \mathbf{\Phi}_u= \{\hat \psi_{1,u}, \hat \psi_{2,u},\dots, \hat \psi_{P',u}\}, \\
 \label{set m up}
  \mathbf{m}_u^\nearrow = \{\hat m_{1,u}^\nearrow , \hat m_{2,u}^\nearrow ,\dots, \hat m_{P',u}^\nearrow \}.
\end{align}
\eqref{set phi up} and \eqref{set m up} can be summarized as following assumption:

% By substituting \eqref{ttd in est set} and \eqref{ps in est set} into \eqref{p-th dft in r antenna appendix}, the DFT-angle sequence
% $R_{c ,g,u}[m,{\bar\psi _{c ,g,u}}]$ can be rewritten as
% \eqref{p-th dft in r antenna with ag}, at the top of this page.
% Here, we assume that ${ R_{c,g,u}}[m,{\bar \psi}_{c,g,u}]$ is equivalent to  the sum of signals from $P'$ distinct paths. In addition, we omit $\tilde{t}$  since it can be treated as a part of the equivalent channel delay.

% To illustrate the relationship of the sweeping angle and the spatial angle of the signal, we make the following assumption:
% By renumbering the elements of $\mathbf{Q}_u$, we can represented the detected angle set as
% \begin{align}
%    \mathbf{Q}_u = \left\{\tilde \psi_{1, u},  \tilde  \psi_{2, u}, \ldots, \tilde{\psi}_{P', u}\right\},
% \end{align}
% where we have $P'=P$ in the ideal noise-free condition.
% To simplify the analysis, in this work we consider $P'=P$.

{\textbf{Assumption 1:}} \emph{ During the angle sweeping phase, the angle detected at the $g$-th slot with $c$-th RF chain, (i.e., $\bar \psi_{c,g, u}\in \mathbf{Q}_u$), is associated with the $p'$-th scattering path with angle  $\psi_{p',u}$ for $p'=1,2,\dots,P'$. } 
{

\begin{figure*}[!htbp]
\vspace{-0.8cm}
{
\begin{align}\label{p-th dft in r antenna with ag}
&{ R_{p', u}}[m] \approx  \sum\limits_{d= 1}^{N_T} 
\sum\limits_{s = 1}^{N_P} { \alpha^g_{p',u} }
e^{-j2\pi ((d-1)(\psi_{p',u}-\bar \psi_{c,g,u})N_P)\left(1 - \left((d-1)(\psi_{p',u}+\bar \psi_{c,g,u})N_P\frac{1}{f_cT_s} + 2{\ell^g_{p',u}}  \right) \frac{\kappa}{f_cT_s} \right)}
% \notag \\
% &\times
 \notag \\
&\times e^{j2\pi\kappa (\ell^g_{p',u})^2}
\frac{\sin\left(-\pi\left(m-M\nu_{p',u} T_s + 2M\kappa{\left({\ell^g_{p',u}} + (d-1)({\psi_{p',u}}+\bar \psi_{c,g,u})N_P\frac{1}{f_cT_s} + {(s-1)\frac{\psi _{p',u}}{f_cT_s}}\right)}\right)\right)}
{M\sin\left(-\frac{\pi}{M}\left(m-M\nu_{p',u} T_s + 2M\kappa{\left({\ell^g_{p',u}} + (d-1)({\psi_{p',u}}+\bar \psi_{c,g,u})N_P\frac{1}{f_cT_s} + {(s-1)\frac{\psi _{p',u}}{f_cT_s}}\right)}\right)\right)}
\notag \\
&\times
 e^{-j\pi \frac{M-1}{M} \left(m-M\nu_{p',u} T_s + 2M\kappa{\left({\ell^g_{p',u}} + (d-1)({\psi_{p',u}}+\bar \psi_{c,g,u})N_P\frac{1}{f_cT_s}  \right)} \right)}
  e^{-j2\pi(s-1)\left({\psi_{p',u}} + \bar \psi_{c,g,u} \right) \left(1+ (M-1)\frac{\kappa \psi_{p',u}}{f_cT_s} \right)},
\end{align}
}
\hrulefill
\end{figure*}

\subsubsection{Analysis of Peak Index}
For ease of explanation, we use the $p'$-th path  detected by the $g$-th TDD slot of the $c$-th RF chain as an example.
By substituting \eqref{ttd in est set} and \eqref{ps in est set} into \eqref{p-th dft in r antenna appendix},
the noise-free sequence related to $p'$-th scattering path can be expressed as \eqref{p-th dft in r antenna with ag}, at the top of this page, where the subscript $c,g,u$ is omitted for simplification.
% The BS can detect the scattering path by configuring sweeping angle ${\bar \psi _{c,g,u}}$ 
% with \eqref{ttd in est set} and \eqref{ps in est set} from set $\mathbf{A}$.
% With {Assumption 1} and
Since the Dirichlet Sinc funtion ${\rm{sinc}} (x)$ attaining its maximum value at $x=0$, 
the relation between ${\hat m _{p',u}^{\nearrow  }}$ and channel parameters can be expressed as
  \begin{align} \label{up chirp max point}
{\hat m _{p',u}^{\nearrow  }} =& -2\kappa M 
\left(\ell^g_{p',u} + \frac{(N_P -1 ) \psi_{c,u}}{2f_cT_s} \right )
+M \nu_{p',u} T_s 
\notag \\ 
& 
 -(N_T-1) ({\bar \psi _{c,g,u}}  +\psi_{p',u}){N_P}  \frac{\kappa M}{f_cT_s} ,
\end{align}
where $\bar \psi_{c,g, u} \in \mathbf{Q}_u$.
Equation \eqref{up chirp max point} illustrates that the peak index ${\hat m _{c,g}^{\nearrow }} $  is determined by delay, Doppler and angle, which are unknown and desired parameters.
Then, BS can derive the channel parameters using the peak index. We will introduce the details in Section III-C.
}

 }
% For unity of notation and ease of exposition, we consider that the parameters of the $p$-th path can be derived with $\hat \psi_{p,u}$ and $\hat m_{p,u}$.

\subsection{Beamforming Receiving for Down-chirp Pilot}
In the beamforming receiving phase,
the user transmits one single down-chirp pilot $\grave C[i]$ at the $(G+1)$-th slot.
% We assume that the angle and Doppler shift are constant during the estimation interval.
Let $\bar{\psi}_{c', G+1, u} $ denote the sweeping angle of the $c'$-th RF chain for down-chirp pilot, where $\cup_{c'=1}^{N_{R}} \bar{\psi}_{c', G+1,u} = -\mathbf{\Phi}_u$.
Then,
% By setting $\bar{\psi}_{c, G+1,u} =-\hat{\psi}_{p,u}$ with $\hat{\psi}_{p,u} \in \mathbf{\Phi}_u$ and  $\cup_{c=1}^{N_{R}} \bar{\psi}_{c, G+1,u} = -\mathbf{\Phi}_u$ , 
the BS can receive down-chirp pilot from $P$ scattering paths simultaneously.
% i.e., we have $\bar{\psi}_{p, G+1} + \psi_{p, u} = 0$. 
Similar to the analysis of up-chirp pilot, the TTD and PS can be optimized by following:
\begin{align}
{t^{G+1}_{c' ,d}} =& \tilde{t} + (d-1) {N _P}{\bar \psi _{c',G+1,u}}/{f_c}, \\
 \Psi^{G+1}_{c' ,d ,s} =& 
 (1-s)  \left ( 1 
 - \frac{\kappa  (M-1)}{T_sf_c} \right){\bar \psi _{c',G+1,u}}.
  \end{align}
The DFT-angle sequence for the down-chirp pilot can be expressed as \eqref{p-th dft in for down}, at the top of { the next} page, 
\begin{figure*}[!htbp]
\vspace{-0.8cm}
\begin{align}\label{p-th dft in for down}
&{ {\grave R}_{c', u}}[m,\bar \psi_{c',G+1,u}] \approx \sum\limits_{p'= 1}^{P'} \sum\limits_{d= 1}^{N_T} 
\sum\limits_{s = 1}^{N_P} {  \alpha^{G+1}_{p',u} }
e^{-j2\pi ((d-1)(\psi_{p',u}+\bar \psi_{c',G+1,u})N_P)\left(1 + \left((d-1)(\psi_{p',u}+\bar \psi_{c',G+1,u})N_P\frac{1}{f_cT_s} + 2{\ell^{G+1}_{p',u}}  \right) \frac{\kappa}{f_cT_s} \right)}
% \notag \\
% &\times
 \notag \\
&\times e^{-j2\pi\kappa (\ell^{G+1}_{p',u})^2}
 e^{\! -j\pi \frac{M\! - \! 1}{M} \left(m \!-M\nu_{p'\!,u} T_s \!- 2M\kappa{\left({\ell^{G+1}_{p',u}} \! + (d-1)({\psi_{p'\!,u}}\!+\bar \psi_{c',G+1,u})N_P\frac{1}{f_cT_s}  \right)} \right)}
  e^{\! -j2\pi(s-1)\left({\psi_{p'\!,u}}\! + \bar \psi_{c',G+1,u} \right) \left(1- (M \!-1)\frac{\kappa \psi_{p',u}}{f_cT_s} \right)}
\notag \\
&\times \!
\frac{\sin\left(\! -\pi \! \left(m\! - \!M\nu_{p',u} T_s \! - \! 2M\kappa{\left({\ell^{G+1}_{p',u}} 
 \! + \! (d \! - \!1)({\psi_{p',u}} \! + \! \bar \psi_{c',G+1,u})N_P\frac{1}{f_cT_s} \! + \! {(s \! - \!1)\frac{\psi _{p',u}}{f_cT_s}}\right)}\right)\right)}
{M \! \sin\left(\! -\frac{\pi}{M} \! \left(\! m \! -\! M\nu_{p'\!,u} T_s \! -\!  2M\kappa{\left({\ell^{G+1}_{p',u}} \! +\!  (\! d\! -\! 1)({\psi_{p'\!,u}}\! +\! \bar \psi_{c',G+1,u})N_P\frac{1}{f_cT_s} \! +\!  {(\! s\! -\! 1\! )\frac{\psi _{p',u}}{f_cT_s}}\right)}\right)\right)}
\!+ \! W_{c',G+1,u}[m].
\end{align}
\hrulefill
\end{figure*}
where $\ell^{G+1}_{p',u}$ is the delay for down chirp and can be expressed as
\begin{align}\label{delay in down}
\ell^{G+1}_{p',u} = \ell^g_{p',u} - \frac{\nu_{p',u}}{f_c} (M+N_{CCP}) (G-g+1).
\end{align} 
We omit the subscript $G+1$ of ${ {\grave R}_{c', u}}$ for simplicity.
For ease of exposition, we assume that the $c'$-th RF chain is align to the $p'$-th detected path for down-chirp case.
The BS can derive the peak index of ${ {\grave R}_{c', u}}[m,\bar \psi_{c',G+1,u}] $ as
 \begin{align}
m_{c',G+1,u}^{\searrow }\! =\! &  \!\!\!\!\!\!\!\mathop{\arg\max}\limits_{m=-M/2,-M/2+1,\dots,M/2-1} \!\!\!\!\!\!\!
\left | { {\grave R}_{c\!,u}}[m,\bar \psi_{c'\!,G+1,u}]
\right |.
 \end{align}
Similar to \eqref{est m} and \eqref{set m up}, peak index set  $\mathbf {m}_{u}^{\searrow }$ is expressed as
% using the Jocabsen estimator similar to \eqref{est m}. 
% The set $\mathbf {m}_{u}^{\searrow }$ can be expressed as
\begin{align}
  \mathbf{m}_u^\searrow = \{\hat m_{1,u}^\searrow , \hat m_{2,u}^\searrow ,\dots, \hat m_{P',u}^\searrow \}.
\end{align}
{
The relation between peak index $\hat{m}_{p', u}^{\searrow }$  and channel parameters can be expressed as
\begin{align}
\label{down index}
\hat m_{p',u}^{\searrow } \!= & 2M\kappa \! \left(\! \ell^{G+1}_{p',u}\! + \! \frac{(N_P \! - \! 1 ) \psi_{p',u}}{2f_cT_s} \! \right ) \! + \! M\nu_{p'\!,u}T_s,
\end{align}
where we assume  $\bar{\psi}_{c', G+1,u} =-\hat{\psi}_{p',u} = -{\psi}_{p',u}$.
}

\subsection{Channel Parameters Computation}

By solving equations \eqref{delay_at_g},
\eqref{up chirp max point}, \eqref{delay in down} and \eqref{down index} under the premise of Assumption 1,
Doppler and delay 
can be respectively derived by 
\begin{align}\label{velocity est}
\hat \nu_{p',u}& = 
 \left(\hat m_{p',u}^{\searrow} \!+\! \hat m_{p',u}^{ \nearrow } \!+ \!
\frac{N_T-1}{f_c T_s} {( \bar \psi_{c,g,u}+ \hat \psi _{p',u} )} N_{P}M\kappa \right)  \notag 
\\
&\times \frac{1}{2MT_s - \frac{2\kappa M(M+N_{CCP})N_{g}}{f_c}},
\end{align}
\begin{align}
\label{delay est}
\hat \ell^{G+1}_{p',u}&= 
 \!- \frac{ { \hat \psi_{p',u}} (N_P-1)}{2f_cT_s}
 \!-\!(N_T\!-\!1)(\bar \psi_{c,g,u} \!+\! \hat \psi_{p',u})\frac{N_P }{4f_cT_s} 
 \notag \\
&+ \! \frac{1}{4M\kappa}(\hat m_{p',u}^{\searrow}  \!-\! \hat m_{p',u}^{\nearrow })\! -\! \frac{ (M\!+\!N_{CCP})N_g \hat \nu_{p',u}}{2f_c},
\end{align}
where $N_g=(G-g+1)$ is the slot deviation. 
%Similar to \eqref{max R} and \eqref{p-th dft in r antenna with ag}, we denote the demodulated pilot in down-chirp case as ${ \mathbf{R}' [{m}, \bar \psi' _{a,g}]}$, which can be obtained by replacing $\kappa$ with $-\kappa$ in \eqref{max R} and \eqref{p-th dft in r antenna with ag}.
By substituting \eqref{down index}, \eqref{velocity est}, \eqref{delay est} into  \eqref{p-th dft in for down},
%$ \rm{R}'_{p,s} [{m}]$, 
we can derive the path gain as
\begin{align} 
 {  \hat \alpha^{G+1}_{p',u} } = & \frac{{ {\grave R}_{c', u}}[{m ^{\searrow}_{c',G+1,u}},\bar \psi_{c',G+1,u}] }{N_S}e^{j2\pi \kappa 
 (\hat \ell^{G+1}_{p',u})^2} \notag \\ 
&\times 
e^{j\pi \frac{M-1}{M} ({m ^{\searrow}_{c',G+1,u}}-M\hat \nu_{p',u} T_s - 2M\kappa{{\hat \ell^{G+1}_{p',u}}})},
\end{align}
where we assume $\bar{\psi}_{c', G+1,u} =-{\psi}_{p',u}$.
%is the down-chirp pilot signal with scanning angle $\bar \psi' _{a,g}$ after the processing of dechirp and DFT. 
%The terms with negligible phase are omitted for simplification.

%Compared to the conventional multi-carrier channel estimation, which requires one DFT, $M$ multiplication and one IDFT operations, our proposed estimation scheme only requires one DFT and $M$ multiplication. 
%Therefore, the computational complexity of channel estimation is reduced.

% \begin{figure}[htbp]
%  \centering
%  \includegraphics[width=75mm]{PDMA.png}
%  \caption{The resource allocation scheme in the delay-Doppler-angle domain.}
%  \label{PDMA}
% \end{figure}
{
\subsection{ Complexity Analysis}

\subsubsection{Algorithm Complexity}  For the processing of channel estimation, the computation be divided into  analog part and digital part. 
The computational complexity of the analog part can be neglected, as it only involves the use of TTD and PS.
The digital part consists of four parts, i.e., dechirp, DFT, path detection, peak searching and parameter computation.
Dechirp is to multiply $N$ samples with the local chirp sequence, with  complexity expressed as $\mathcal{O} (M)$.
The computation complexity of DFT can be expressed as $\mathcal{O} (M \log (M))$.
Path detection determines whether the magnitude of $N_A$  angle-domain samples exceeds a predefined threshold, which exhibits a complexity of  $\mathcal{O} (N_A)$.
The peak searching is to find the maximum magnitude from $M$ samples with the complexity expressed as $\mathcal{O} (M)$.
The Jocabsen estimator involves only simple addition and division operations, and its computational complexity is negligible.
The channel parameters computation relies exclusively  on elementary arithmetic, thereby rendering the computational complexity negligible.
Thus, the overall complexity is dominated by $\mathcal{O} (N_{A}M \log (M) +N_A+ 3M)$.
{In contrast, we list the complexities of the channel estimation algorithms  in TABLE I  based on grid searching \cite{squint_2} and   sparse Bayesian learning (SBL)  \cite{pc-sbl}, where $I$ represents the number of iterations.
}
\begin{table}[htbp]\label{complexity}
\centering
    \caption{Complexity of different algorithms}
\begin{tabular}{|l|c|r|}
  \hline
  Alogrithm & Computation complexity\\
  \hline
  Grid searching-based \cite{squint_2} & $\mathcal{O}(N_A M \log N_A M +  P N_A M + N_A M ) $  
  \\
  % \text{\cite{squint_2}} &$+ N_A M )$\\
  \hline
  % OMP-based \cite{omp-based} & $\mathcal{O}(P I N_A^{2} + P^{4} \ln\left(\frac{N_A}{P I}\right)+ P I N_A G)$\\   
  % \hline
  SBL-based \cite{pc-sbl} & $\mathcal{O}\left(   16I(M N_{RF})^2 MN_A\right)$\\
  \hline
  Proposed & $\mathcal{O} (N_{A}M \log (M) +N_A+ 3M)$\\
  \hline
\end{tabular}  
\end{table}

\subsubsection{Pilot Cost} Considering the pilot structure in  \figurename{ \ref{transmission_model}}, with respect to each user, BS is required to transmit $G+1$ chirp-based pilots, each with $M+N_{CCP}$ sampling points. Thus, the pilot costs can be expressed as $(G+1)(M+N_{CCP})$.

}

\begin{figure*}
\vspace{-0.8cm}
  \begin{align}\label{H_p_l from dd domain to tf domainRe}
&\tilde{h}_{k, \!\ell}^{u,u^{\prime}} \! \left[k^{\prime},\ell^{\prime}\right] 
\!=\!
%\notag \\ & 
 \sum_{p=1}^{P}  \!
\sum_{c=1}^{N_{R}} \!
\frac{\sqrt{\rho_{c}}}{NM} 
{e^{j2\pi \nu_{p,u}\frac{\ell'}{M} }}
\sum_{n=0}^{N-1} \sum_{m=0}^{M-1} \!
%e^{j2\pi\left(\frac{nk}{N} -\frac{m\ell}{M}  \right)}
%e^{j2\pi\left(\frac{nk^{\prime}}{N} - \frac{m \ell^{\prime}}{M}\right)}
e^{j2\pi\left(\frac{n(k^{\prime}-k)}{N} - \frac{m(\ell^{\prime}-\ell)}{M}\right)} 
\alpha_{p,u} 
 e^{-j2\pi\frac{m\ell_{p,u}}{M}} e^{j2\pi n\left(\frac{k_{p,u}}{N} + \frac{m}{\mu_{p,u}}\right)}
 \notag \\
 &\times
 \sum_{d=1}^{N_{T}} \sum_{s=1}^{N_{P}} 
e^{-j2\pi\left(\left((d-1)N_{P}+s-1\right)\psi_{p,u} + t_{c,d}f_{c}\right)\left(1 + \frac{m\Delta f}{f_{c}}\right)} 
% \notag \\
%& \times 
e^{j2\pi\Psi_{c,d,s}} 
e^{j2\pi\nu_{p,u}\frac{\left( \tilde{\ell}_{c,d} + \ell_{p,u} + \left((d-1)N_{P}+s-1\right)\frac{\psi_{p,u}}{f_{c}T_{s}}\right)T}{M}}
% \notag \\
% & 
\end{align}
 %\hrulefill
\end{figure*}

\section{Proposed Downlink Hybrid Precoding Design}
{
The doubly squint effect seriously destroys the sparsity of the DD-domain channel, which increases the complexity of the direct precoding in the DD domain.
In this work, we consider a new hybrid precoding scheme using the channel parameters estimated in Section III.
Equation \eqref{H_p_l from dd domain to tf domain} can be rewritten as \eqref{H_p_l from dd domain to tf domainRe}, at the top of the next page.
From \eqref{H_p_l from dd domain to tf domainRe}, the effect of Doubly squint effect can be decomposed into phase shifts in DD domain and TF domain.
Therefore, it is feasible to to compensate the phase shifts in DD and TF domains  to realize precoding.
Instead of direct precoding in the DD domain with convolution  \cite{dd_precoding},
we consider a hybrid precoding with digital part compensating the phase deviation in both DD domain and TF domain.
}
% By utilizing parameters estimated in Section III, 
% we can optimize our proposed hybrid precoding and reduce the doubly-squint effect.
% $\hat \tau^{G+1}_{p',u} = T_s\hat \ell^{G+1}_{p',u}$
% % $
% % \hat \nu_{p',u},
% % \hat{\psi}_{p',u}, \hat{\alpha}_{p',u}$, 
% \eqref{delay_at_g} and the channel reciprocity in TDD mode, we can obtain the channel parameters when downlink transmission. 
To simplify notation, we denote $\left\{\hat \tau_{p',u}, \hat \nu_{p',u}, \hat{\psi}_{p',u}, \hat{\alpha}_{p',u} \right\}$ as the channel parameter used in our proposed hybrid precoding. The input-output relationship in \eqref{IO_relation} can be rewritten as
\begin{align}\label{io_dd_domain}
&y_u[k,\ell]=
\frac{1}{NM} h_{k,\ell}^{u,u}[k,\ell] x_{u}[k,\ell]
\notag \\
&
+ \! \frac{1}{NM} \! \sum_{k'=0}^{N-1}  \sum_{\ell'=0}^{M-1}  \sum_{u'=0,u'\neq u}^{U}  \!\! h_{k,\ell}^{u,u'}\! [k'\!,\ell'] x_{u'}[k'\!,\ell']
\notag \\
&
+ \! \frac{1}{NM} \! \sum_{k'=0,k'\neq k}^{N-1}  \sum_{\ell'=0,\ell'\neq \ell}^{M-1} 
\!\! h_{k,\ell}^{u,u}[k'\!,\ell'] x_{u}[k'\!,\ell']
%+ \not y_s[k,l]
\!+ \!w_u[k,\ell],
\end{align}
where the first item corresponds to the desired signal, 
the second is inter-user interference 
the third refers to ISI,
and the forth is AWGN in the DD domain with variance $\sigma^2$.
 {Therefore, the achievable rate per DD grid can be represented as
\begin{align}\label{achievabla rate}
{C^u_{k,\ell}} = {\log _2}\left( 1 + \text{SINR}_{k,\ell} \right),
\end{align}
where $\text{SINR}_{k,\ell}$ is the signal to interference plus noise ratio (SINR) of the DD grid $x_{u}[k,\ell]$, which can be represented as \eqref{SINR}, at the top of this page. 
In order to maximize the achievable rate, we then optimize our proposed precoding method by maximizing the SINR in the following analysis.}

\begin{figure*}[!htbp]
\vspace{-8mm}
\begin{align}\label{SINR}
{SINR_{k,\ell}} =  { \frac{{{{\left| { h_{k,\ell}^{u,u}[k,\ell] } \right|}^2}}}{{
\sum_{k'=0,k'\neq k}^{N-1}  \sum_{\ell'=0,\ell'\neq \ell}^{M-1} \left|h_{k,\ell}^{u,u}[k',\ell']\right|^2
+
\sum_{k'=0}^{N-1}  \sum_{\ell'=0}^{M-1}  \sum_{u'\neq u}^{U}  \left| h_{k,\ell}^{u,u'}[k',\ell']  \right|^2
+
{\sigma  ^2}}}} ,
\end{align}
 \hrulefill
\end{figure*}

From \eqref{H_p_l from dd domain to tf domain}, the beam squint effect can be mitigate by configuring TTDs and PSs  as
\begin{align}\label{ttd in precoding}
 { t_{c ,d}} = - \kappa {N _P}{\hat \psi _{p',u}}/{f_c},
 \\
%\end{align}
%\begin{align}\label{ttd in precoding}
\label{ps in precoding}
{ \Psi_{c ,d,s}} = (s-1){\hat \psi _{p',u}}.
\end{align}
\eqref{ttd in precoding} and \eqref{ps in precoding} illustrate that each RF chain is aligned to a specific angle \cite{squint_4}.
%  This feature limits the flexibility of precoding. 
% Fortunately, the PDMA scheme \cite{OTFS_7} is particularly adept at fulfilling this requirement, as it categorizes users based on the overlap of their scattering paths in the angular domain.
This requirement is well-suited for the PDMA scheme \cite{OTFS_7}, as it schedules user access  based on the overlap of their scattering paths in the angle domain.
% as shown in \figurename{ \ref{PDMA}}.
By using the scheduling strategy of PDMA,  the inter-user interference is negligible   when the number of antenna is large enough.
% Then, 
% ${h}_{k,\ell}^{u,u^{\prime}}\left[k^{\prime},\ell^{\prime}\right]$ can be simplified as ${h}_{k,\ell}^{u,u^{\prime}}\left[k^{\prime},\ell^{\prime}\right] = 0$ when $u \neq u' $.
% \begin{align}\label{h uu2}
% {h}_{k,\ell}^{u,u^{\prime}}\left[k^{\prime},\ell^{\prime}\right] = 
% \left\{ \begin{gathered}
%   0,u \neq u'  \hfill  \\
% {h}_{k,\ell}^{u}\left[k^{\prime},\ell^{\prime}\right],u=u' \hfill \\ 
% \end{gathered}  \right.,
% \end{align}
Thus, only the case of $u'=u$ is considered in \eqref{io_dd_domain}.

In PDMA scheme, each RF chain can transmit signals to a unique scattering path with a specific user.
We assume that the signal related to the $p'$-th path is precoded by the $c'$-th RF, where $c'$ is mapping by  $c' = \mathcal{M}(u,p') $.
Considering that  the DD-domain digital precoding $D_{c',u}[k',\ell']$ contains the mapping relationship between RF chain and scattering path, $D_{c',u}[k',\ell']$ can be simplified to 
% Therefore, the DD-domain digital precoding $D_{c',u'}[k',\ell']$, which  contains the mapping relationship between RF chain and scattering path, is expressed as 
\begin{align}
{{D}}_{c',u}[k',\ell'] =
\left\{ \begin{gathered}
  0, c'\neq \mathcal{M}(u,p')  \hfill \\
\bar {{D}}_{p',u}[k',\ell'] , c' = \mathcal{M}(u,p') \hfill \\ 
\end{gathered}  \right.,
\end{align}
where $\bar {{D}}_{p',u}[k',\ell']$ is used to compensate the
 phase deviation that related to $k'$ and $\ell'$. From \eqref{H_p_l from dd domain to tf domain} $\bar {{D}}_{p',u}[k',\ell']$ is given by
\begin{align}\label{DD domain precoding}
\bar {{D}}_{p',u}[k',\ell'] =& 
{e^{ - j2\pi {{\hat \nu}_{p',u}}\frac{{\ell' T}}{M}}} \times \left\{ \begin{gathered}
  1,\ell' \in \mathcal{L}_{ICI}^{p',\;u} \hfill \\
  {e^{j2\pi \frac{k}{N}}},\ell' \in \mathcal{L}_{ISI}^{\;p',u} \hfill \\ 
\end{gathered}  \right..
\end{align}
% is used to compensate the
%  phase deviation that related to $k'$ and $\ell'$ in \eqref{H_p_l from dd domain to tf domain}. From
% \eqref{H_p_l from dd domain to tf domain}, 
% $\bar D_{p',u'}[k',\ell']$ can be expressed as
By using \eqref{DD domain precoding}, ${h}_{k,\ell}^{u,u}\left[k',\ell'\right]$ can be expressed as
 \eqref{h simple without dd-precoding}.
 % by substituting \eqref{ttd in precoding}, \eqref{ps in precoding} and \eqref{DD domain precoding} into \eqref{H_simple}.
Here, we define $\bar \rho_{p',u} =  \rho_{c'}$ to unify notation.
Without loss of generality, we consider $e^{-j2\pi \left(s-1\right)\psi_{p',u} \left( \frac{m\Delta f - \nu_{p',u}}{f_{c}}\right)} \approx 1$ when $N_T$ is large enough.
% (i.e., $N_P$ is small enough).
% $\sum_{s=1}^{N_{P}} e^{-j2\pi \left(s-1\right)\psi_{p',u'} \left( \frac{m\Delta f - \nu_{p',u'}}{f_{c}}\right)}$ is the error term, which is approximately equal to $N_P$ if $N_T$ is large enough.
\begin{figure*}[!htbp]
\vspace{-0.8cm}
    \begin{align}\label{h simple without dd-precoding}
&{h}_{k,\ell}^{u,u}\left[k^{\prime},\ell^{\prime}\right]  =  \sum_{n=0}^{N-1} \sum_{m=0}^{M-1} 
\sum_{p'=1}^{P'} 
\sum_{s=1}^{N_{P}} e^{-j2\pi \left(s-1\right)\psi_{p',u} \left( \frac{m\Delta f - \nu_{p',u}}{f_{c}}\right)}
\frac{N_T\sqrt{\bar \rho_{p',u}}}{NM} \alpha_{p',u} e^{j2\pi\left(\frac{n(k^{\prime}-k)}{N} - \frac{m(\ell^{\prime}-\ell)}{M}\right)}
\notag \\
&\times e^{j2\pi\nu_{p',u}\frac{ \ell_{p',u}T}{M}} 
\left\{ \begin{array}{l} e^{-j2\pi\frac{m\ell_{p',u}}{M}} e^{j2\pi n\left(\frac{k_{p',u}}{N} + \frac{m}{\mu_{p',u}}\right)}, \ell^{\prime} \in \mathcal{L}_{ICI}^{p',u}  \\
 e^{-j2\pi\frac{m\ell_{p',u}}{M}} e^{j2\pi\left(n-1\right)\left(\frac{k_{p',u}}{N} + \frac{m}{\mu_{p',u}}\right)}, \ell^{\prime} \in \mathcal{L}_{ISI}^{p',u} \end{array} \right.
\end{align}
\hrulefill
\end{figure*}
% For the sake of illustration, we change the subscript of $\rho$ from $c$ to $p'$ since we assume that the signal of the $c$-th RF chain maps to the $p'$-th path. 
Then, given $c' = \mathcal{M}(u,p') $, the remaining phase shift associated with $n$ and $m$ can be compensated if TF precoding $B_{c'}[n,m]$ is given by
\begin{align}
\label{BTF}
{{B}}_{c'}[n,m] =&
\frac{\hat \alpha_{p',u} ^*}{\left| \hat\alpha_{p',u}  
 \right|}
e^{-j2\pi \hat \nu_{p',u}\bar{\ell}_{u}T_s}  {e^{j2\pi \frac{m}{M}({{\hat \ell }_{p',u} + \bar {\ell}_{u}}
 )}}
 \notag \\
 &\times
 {e^{ - j2\pi n
 \left(
 \frac{{{{\hat k}_{p',u}}}}{N} +\frac{m}{\hat \mu _{p',u}
 } \right)}
 },
\end{align}
where ${\hat k}_{p',u}=\hat{\nu }_{p',u}NT$,
${\bar \ell}_{u}$ is a constant and it is used to ensure
that both index sets of ICI and ISI are suitable for all antennas. 
% To unify the notation, we define $\bar{{B}}_{p',u'}[n,m]={{B}}_{c'}[n,m]$.
The index sets for ICI and ISI are formulated as  \eqref{ICI index set afp} and \eqref{ISI index set afp}, respectively.
Here $N_P$ and $\bar \ell_{u}$ are subject to the constraint as \eqref{st}
% \begin{align}\label{st t}
% \text{s.t.} \quad 0 < -  \frac{{{(s-1) \hat \psi _{p',u'}}}}{{{f_c}{T_s}}} + \frac{M}{{{\hat \mu _{p',u'}}}} + \bar {\ell}_{u'} < 1 , 
% %\quad \forall c , \hat \mu_{p,s},
% \end{align}
for all $1\le s \le N_P$ and $ \left | \frac{1}{\hat \mu_{p',u}} \right| < \frac{\nu_{max}}{f_c} $. 
% Here, $\nu_{max}$ is the maximum Doppler shift.

\emph{Proof:}  Appendix \ref{proof of bd after precoding}.

{
In order to derive the power factor $\sqrt {{\rho _{c'}}}$, we assume that the phase deviations introduced by different paths are compensated by the proposed hybrid precoding.
Then, the power factor can be derived by maximum ratio combining (MRC).
The optimal achievable rate per data symbol can be expressed as
% \eqref{achievabla rate} can be rewritten as
% the achievable rate can be written as
%We then optimize the power factor $\sqrt {{\rho _{c}}} $ by maximizing the achievable rate, which 
%%The achievable rate for the $u$-th user
% can be expressed as
\begin{align}\label{c simple}
{C^{u}_{k,\ell}} = {\log _2}\left( {1 + \frac{{{{\left| {\sum\limits_{p' = 1}^{{P'}} {N_A\sqrt {{\bar \rho _{p',u}}} \left| { { \alpha} _{p',u}} \right|} } \right|}^2}}}{{
{\sigma  ^2}}}} \right).
\end{align}
}
%where $\sigma  ^2$ represents the power of AWGN.
%In order to maximize the achievable rate, we adjust the power factor by the maximum ratio combining  as
Let $E_{u}$  denote  total power of transmitted signal allocated to the $u$-th user.
By using MRC, the optimal power factor $\sqrt {{\rho _{c'}}}$ can be derived as
%Consequently, the optimal power factor for the $p$-th user can be expressed as
%the fourth user can be expressed as
\begin{align}\label{power allocation}
\sqrt {{\rho _{c'}}}  = \sqrt {{\bar \rho _{p',u}}}  = \sqrt {\frac{{{E_{u}}}}{{MNN_A}}} \frac{{\left| {{{{\hat { \alpha} } }_{p',u}}} \right|}}{{\sqrt {\sum\limits_{p' = 1}^{{P}} {{{\left| {{{{\hat { \alpha}} }_{p',u}}} \right|}^2}} } }}.
%p = 1,2,...,{P}.
\end{align}

\section{Simulation Results}

%In this section, we evaluate the performance of our proposed algorithm for channel estimation and hybrid precoding. The typical value of relevant simulation parameters is provided in \tablename{ \ref{sim para}}. 
%We use normalized mean square error (NMSE) as the performance indicators of uplink channel parameter estimation, which can be expressed as \eqref{nmse}, where $\hat{x}$ consists of $\alpha, \tau, v$ and $\psi$, The Signal-to-noise ratio of the uplink received signal is expressed as ${SNR}^{ul} = 10\log (\sigma _y^2/{\sigma ^2})$.

In this section, the efficacy of the proposed method for channel estimation and hybrid precoding are evaluated. 
Unless otherwise stated, our main simulation parameters are given in \tablename{ \ref{sim para}}. 
The antenna spacing of BS is set to ${\lambda_c}/{2}$.
The power of the multipath channel gain is normalized, i.e. $\sum\nolimits_{p = 1}^{{P}} {{{\left| {{{\bar \alpha }_{p,u}}} \right|}^2}}  = 1$.
%We use the normalized mean square error (NMSE) as the indicator for uplink channel parameter estimation, which can be expressed as \eqref{nmse},
%where $\hat{x}$ consists of $\alpha$, $\tau$, $v$, and $\psi$.
The signal-to-noise ratio (SNR) is expressed as $\text{SNR} = 10\log (\sigma _y^2/{\sigma _n^2})$, where $\sigma _y^2$ and ${\sigma _n^2}$ are the power of the received signal without noise and the power of AWGN, respectively.
%By adopting the Monte Carlo method, 
The performance metric of channel estimation is the normalized mean square error (NMSE) as
%, which is mathematically expressed as
%\begin{equation}
%\text{SNR}^{ul} = 10 \log \left(\frac{\sigma_y^2}{\sigma^2}\right)
%\end{equation}
\begin{align}\label{nmse}
{\mathbf{NMSE_x}} = \mathbb{E} \left \{ \frac{{{{\left\| {{{\hat {\mathbf x} - \mathbf x}}} \right\|}^2}}}{{{{\left\| \mathbf{{x}} \right\|}^2}}} \right\}  , {\mathbf{x}}=\bm{\alpha}_u, \bm{ \tau}_u, \bm{ \nu}_u,
\end{align}
where $\hat {\mathbf x}$ represents the estimation of ${\mathbf x}$, with its $p$-th item characterized by $\alpha_{p,u}$, $\tau_{p,u}$ and $\nu_{p,u}$. The achievable rate is the performance metric for theproposed hybrid precoding.

\begin{table}[t]
    \centering
    \caption{Simulation Settings}
    \label{sim para}
    \begin{tabular}{lcccrrr}
    \toprule
        \textbf{Parameters} & \textbf{Value}  \\ \midrule
        Carrier frequency & 30 GHz  \\ 
        Bandwith & 1 GHz   \\
        Number of subcarrier & 2048 \\
        Subcarrier spacing & 500 kHz \\
        Number of path & 4 \\
        propagation delay & [0, 10] $Ts$ \\
        UE speed  & 250 km/h\\
        DoA of scattering path & $[-90^{\circ},90^{\circ}]$\\
        Number of TTD line & 8\\
        Number of RF chain & 4\\
        SNR & 15 dB\\
        OTFS frame  $M\times N$ & 2048 $\times$ 128\\
    \bottomrule
    \end{tabular}
\end{table}

%\begin{figure}[htbp]
%  \centering
%  %\label{DDA channel}
%  % Requires \usepackage{graphicx} and \usepackage{subfigure}
%  \subfigure[]
%  {\label{new_patterns2}
%    \begin{minipage}{0.22\textwidth} % 设置子图宽度为页面宽度的45%
%    \centering
%    \includegraphics[width=\linewidth]{angle_gain_r128_fc60_fs2.png} % 子图宽度设置为行宽
%    \end{minipage}
%  }
%  \hfill % 填充空白，分隔两个子图
%  \subfigure[]
%  {\label{new_patterns1}
%    \begin{minipage}{0.22\textwidth} % 设置子图宽度为页面宽度的45%
%    \centering
%    % Requires \usepackage{graphicx}
%    \includegraphics[width=\linewidth]{angle_gain_r16_fc60_fs2.png} % 子图宽度设置为行宽
%    \end{minipage}
%  }
%  \caption{array gain of TTD-PS net  (a) number of antenna r=128, number of TTD K=1; (b) number of antenna r=128, number of TTD K=8.}
%   \label{angle gain}
%\end{figure}

\begin{figure}[htbp]
 \centering
 \includegraphics[width=85mm]{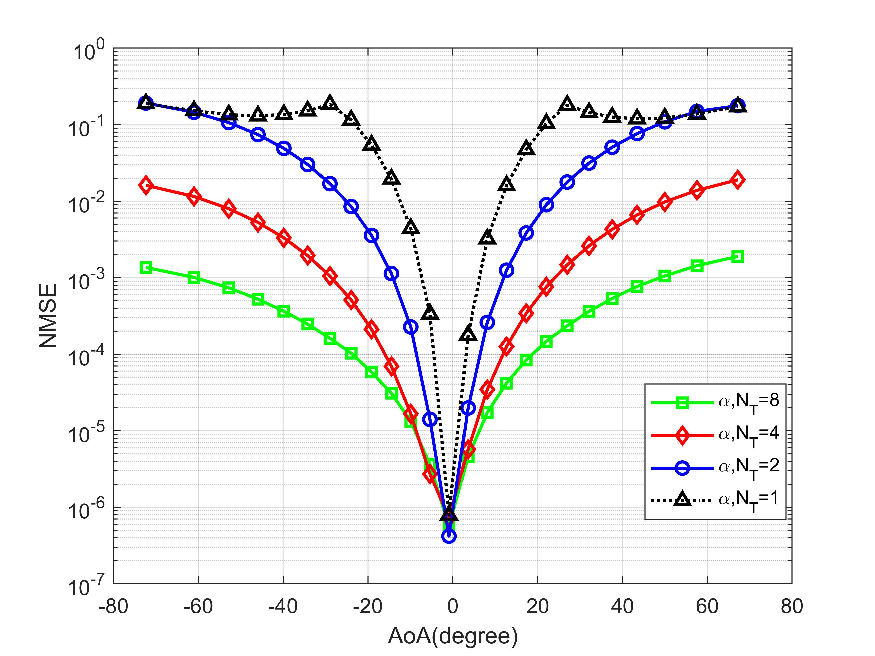}
 \caption{The NMSE of $\bm \alpha_u$ vs AoA at different numbers of TTD.}
 \label{angle nmse}
\end{figure}

\figurename{ \ref{angle nmse}} illustrates the
NMSE performance for path gain $\bm \alpha_u$ at different TTD number, where the conditions with the TTD  numbers $N_T$ of 8, 4, 2 and 1 are considered.
The number of scattering path is fixed as 1.
As shown in \figurename{ \ref{angle nmse}},
the NMSE decreases with the TTD number $N_T$ increases.
This is because the sum over Dirichlet Sinc function in \eqref{p-th dft in for down} has deviation item $ \frac{2M\kappa \psi_{p,u}}{f_c T_s}(s-1)$, for $s=1,2,\cdots,N_P$.
Considering $N_P=N_A/N_T$, the larger values of $N_T$ lead to smaller values of $N_P$, which decrease the impact of the deviation item.
In addition, the NMSE increases when the AoA deviates from $0^{\circ}$.
This is because the deviation item $ \frac{2M\kappa \psi_{p,u}}{f_c T_s}(s-1)$ is also a function of angle $\psi_{p,u}$, the larger $\psi_{p,u}$ leads to a higher deviation.

\begin{figure}[htbp]
 \centering
 \includegraphics[width=85mm]{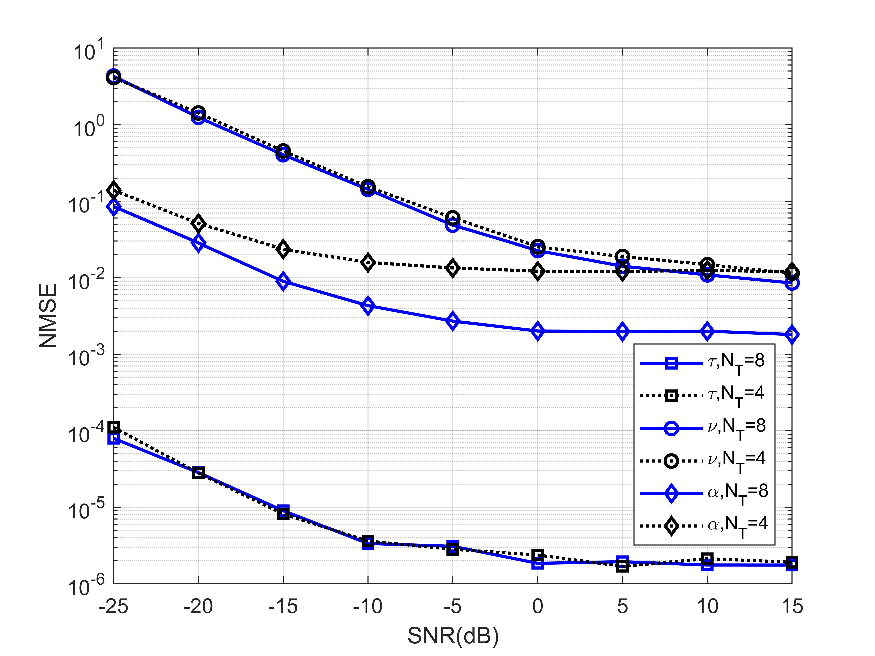}
 \caption{The NMSE of $\bm \tau_u, \bm \nu_u, \bm \alpha_u$, vs SNR,  slot deviation $N_g = 10$.}
 \label{NMSE_figs_para}
\end{figure}
\figurename{ \ref{NMSE_figs_para}} shows the NMSE performance of $\bm \tau_u, \bm \nu_u, \bm \alpha_u$ at different SNR.
It compares two conditions: $N_T=8$ and $N_T=4$ .
The NMSE of $\bm \tau_u$, $\bm \nu_u$, $\bm \alpha_u$ is observed to decreases with increasing SNR for both conditions.
The increase of TTD number has a negligible effect on the NMSE for both $\bm \tau_u$ and $\bm \nu_u$.
However, the NMSE of $\bm \alpha_u$ is notably lower when $N_T=8$ compared to  $N_T=4$.
This discrepancy arises because the estimation of $\bm \tau_u$ and $\bm \nu_u$ relies on the peak index of ${ R}_{c,u}[m,\bar \psi_{c,g,u}] $ and ${ {\grave R}_{c', u}}[m,\bar \psi_{c',G+1,u}]$, which is less sensitive to beam squint effect.
% In contrast, 
The estimation of path gain $\bm \alpha_u$ requires the maximum value of ${ {\grave R}_{c', u}}[m,\bar \psi_{c',G+1,u}]$, which is more sensitive to beam squint effect.

\begin{figure}[htbp]
 \centering
 \includegraphics[width=75mm]{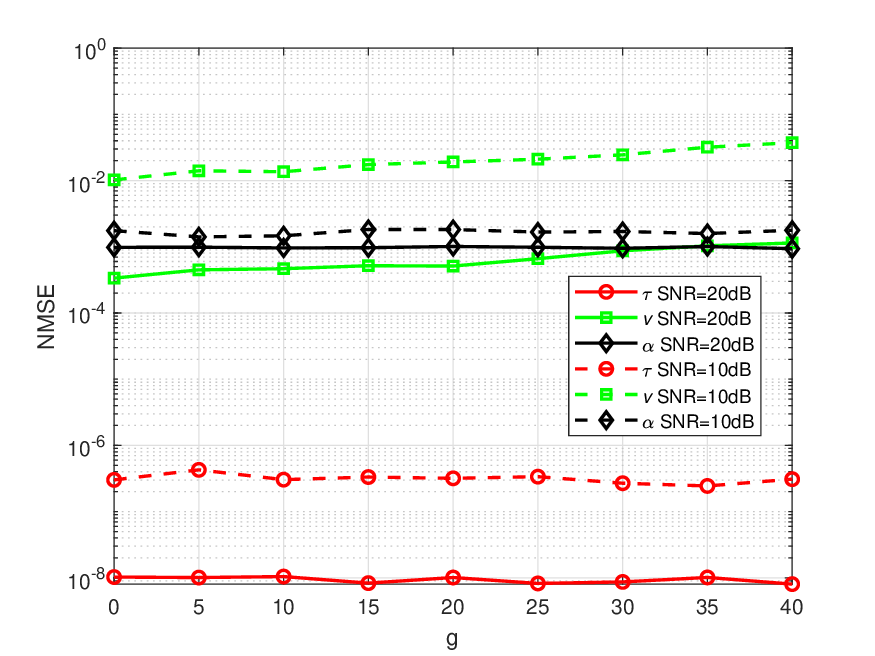}
 \caption{The NMSE versus interval of reception $g$.}
 \label{NMSE vs g}
\end{figure}
{ 
\figurename{ \ref{NMSE vs g}} illustrates the impact of the interval of reception $g$ on the performance of NMSE.
We consider that the beam squint effect are completely eliminated.
From \figurename{ \ref{NMSE vs g}},the NMSE of Doppler estimation decreases as the interval between the up-chirp and down-chirp signals increases.
This trend is attributed to the fact that a larger interval leads to more significant variation in delay, thereby enhancing the ability to resolve delay variations, which resulting the Doppler shift.
Meanwhile, the NMSE of delay estimation is not affected by the change of $g$.
This is due to the high robustness of our super-resolution algorithm to small delay variations.
Additionally, the NMSE of gain remains stable as $g$ varies. 
This stability is beacause the path gain is more sensitive to delay variation rather than Doppler variation. 
When the accuracy of delay estimation is constant and the accuracy of Doppler estimation is slightly changed, the accuracy of path gain estimation remains almost constant.
}

\begin{figure}[htbp]
 \centering
 \includegraphics[width=85mm]{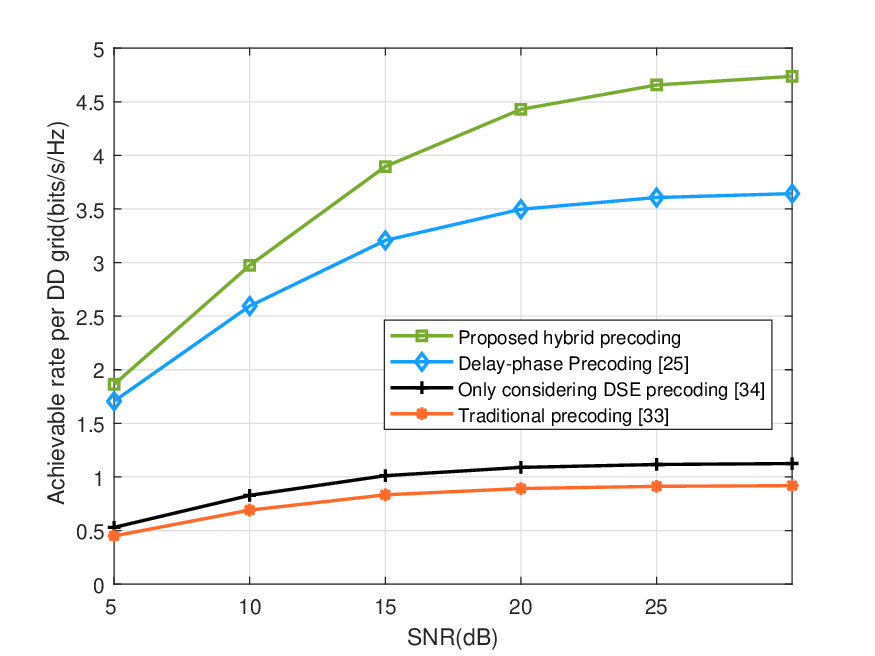}
 \caption{ Achievable rate  per DD grid versus  SNR.}
 \label{NMSE_figs_SNR}
\end{figure}
The comparative analysis of the achievable rate { per DD grid} between the proposed hybrid precoding method and existing methods is depicted in \figurename{ \ref{NMSE_figs_SNR}}.
The alternative precoding schemes include only considering Doppler squint precoding \cite{squint_3}, delay-phase precoding \cite{squint_4} and traditional precoding \cite{OTFS_7}. 
 It is observed that the only considering Doppler squint precoding and traditional precoding  exhibit lower achievable rates.
 This decline is  primarily due to the beam squint effect, which significantly diminishes the array gain in wideband and large-scale antenna systems. 
Delay-phase precoding optimizes the array gain by compensating for the beam squint effect and has suboptimal performance.
Our proposed precoding outperforms the others by effectively mitigating both the beam squint and Doppler squint effects, especially at high SNR conditions.

% achievable rate for four precoding case,
%i.e., i) proposed hybrid precoding, ii) No BeamSquint precoding [], iii) No Doppler Squint precoding [], and iv) traditional precoding. 
%As expected, the traditional precoding scheme maintains the lowest performance due to not considering the beam dispersion and Doppler dispersion issues. As the 

%shows the NMSE performance for $\tau$ and $v$ at different number of TTD elements, where two conditions, i.e., the SNR of -20 dB, -10 dB, 0 dB are consideration. 
%%$K \in \{ 2,4,8,16,32,64,128\}$ are taken into consideration. 
%As shown in \figurename{ \ref{NMSE_figs_comp}}, the NMSE decreases with $N_P$ decreased. This is because the beam squint effect is attenuated. When $K=2$, NMSE of doppler is more than $0$ and parameter estimation could not be performed.
%When the number of TTD elements $K$ exceeds 8, the uplink channel estimation error tends to stabilize, indicating a diminishing return on further increasing the number of delay lines. Consequently, the optimal configuration for the number of true delay lines is established at 8. This selection not only yields a lower channel estimation error but also incurs a reduced hardware cost due to the minimized requirement for additional delay lines.

\begin{figure}[htbp]
 \centering
 \includegraphics[width=85mm]{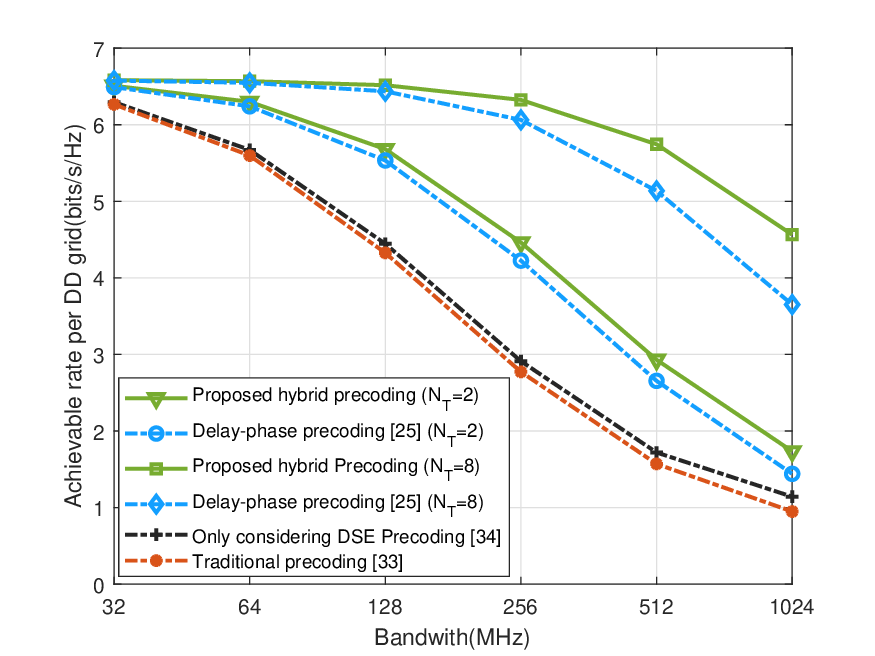}
 \caption{Achievable rate per DD grid versus bandwith.}
 \label{Cq vs bandwith_figs}
\end{figure}
\figurename{ \ref{Cq vs bandwith_figs}} presents a comparison of the average achievable rate { per DD grid versus the bandwith} for different precoding schemes at a SNR of 20 dB.
It is evident that the performance of only considering Doppler squint precoding and traditional precoding schemes exhibit a marked degradation at higher bandwidths.
{
This is due to the fact that given a fixed subcarrier interval and symbol number $N$, an increase in bandwidth leads to an increase in the number of higher subcarriers. 
Thus resulting in a larger $M\times N$ causing a larger Doppler squint effect.
In addition, the larger bandwidth leads to more severe beam squint effect.
The enhancement of both Doppler squint effect and beam squint effect lead to a lower achievable rate as the bandwidth increases.
}
Conversely, both delay-phase precoding and our proposed precoding demonstrate an enhancement in achievable rate performance with an increasing number of  number of TTDs.
Notably, the proposed precoding method demonstrates superior performance due to its consideration of both beam squint and Doppler squint effects.
In addition, our analysis reveals that the proposed precoding scheme substantially enhances the achievable rate compared to delay-phase precoding, particularly when a large number of TDD elements are employed. However, this enhancement is less pronounced when the count of TDD elements is low.

%\begin{figure}[htbp]
% \centering
% \includegraphics[width=85mm]{Cq_bandwith.png}
% \caption{The performance of precoding vs bandwith.}
% \label{Cq vs bandwith_figs2}
%\end{figure}
%\figurename{ \ref{Cq vs bandwith_figs2}} illustrates the achievable rate performance versus bandwith, where number of TD elements K = 1, 8, 16, 32, 128, are considered. When K=128, each antenna is individually connected to a TTD and RF chain, so the effects of beam and Doppler dispersion are completely eliminated.
%It has been found that when the bandwidth is small, less TTD can approach the ideal rate because the influence of beam dispersion is not obvious.
%When the bandwidth is large, the gain decreases due to the decrease of the array gain.

\begin{figure}[htbp]
 \centering
 \includegraphics[width=85mm]{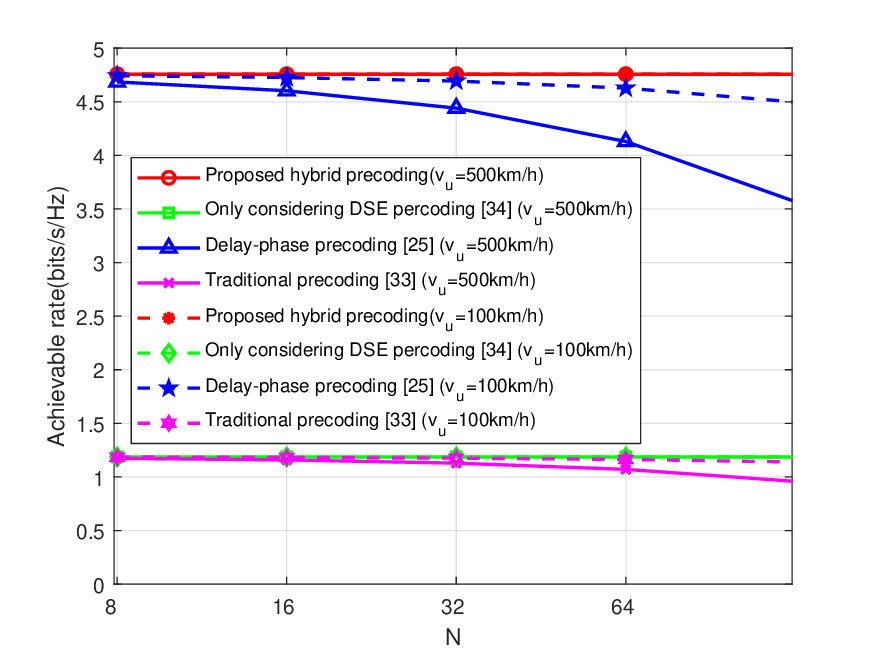}
 \caption{ Achievable rate per DD grid versus N.}
 \label{N vs c}
\end{figure}
\figurename{ \ref{N vs c} }presents a comparison of the average achievable rates {per DD grid} for different percoding schemes and user velocities, where the SNR is 20 dB. 
The parameter $N$ is associated with the duration of the OTFS frame. 
It is observed that precoding schemes that do not account for Doppler squint effect such as the Delay-phase precoding and traditional precoding,  exhibit a decline in achievable rate as $N$ increases, particularly at high velocities.
This degradation is attributed to the Doppler squint effect, which is evident through the factor ${e^{\frac{j2\pi mn}{\mu _{p,u}}}}$ in \eqref{H_p_l from dd domain to tf domain}, leading to a phase deviation among different subcarriers and can be accumulated as $N$ increases.
In contrast, the performance of only considering Doppler squint precoding and the proposed precoding remains constant for all conditions.
It illustrates that the effect of Doppler squint is eliminated.
Moreover, as illustrated in \figurename{ \ref{N vs c}}, the beam squint effect tends to have a greater impact compared to the Doppler squint effect.

\begin{figure}[htbp]
 \centering
 \includegraphics[width=75mm]{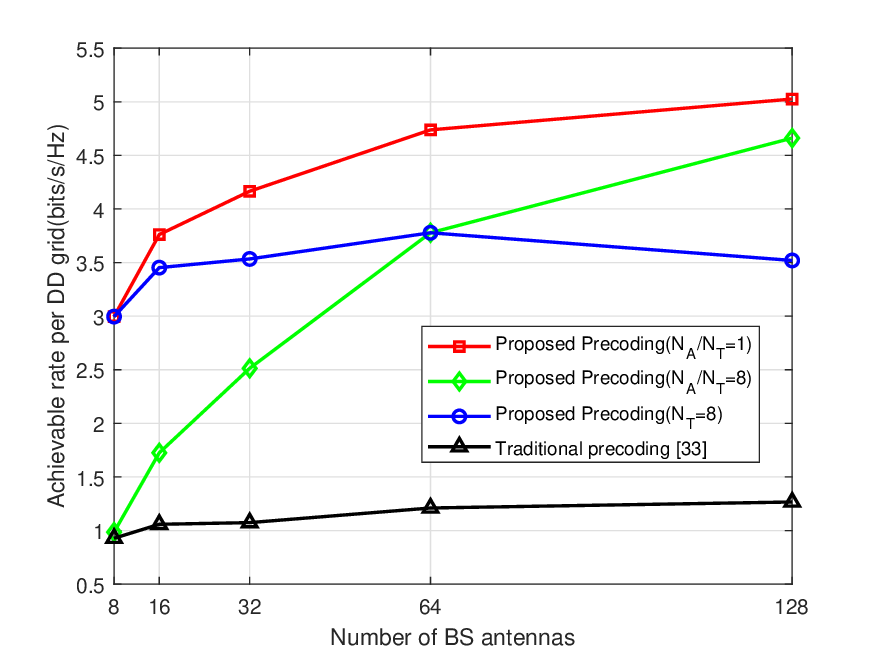}
 \caption{Achievable rate per DD grid versus the number of BS antenna.}
 \label{rate vs antenna}
\end{figure}
{
\figurename{ \ref{rate vs antenna}}
  illustrates the achievable rate per DD grid versus the number of BS antennas. 
When $ N_A /N_T = 1$, the proposed precoding scheme can effectively eliminate both Doppler squint and beam squint, resulting in the maximum achievable rate.
For the case where the number of TTD elements $N_T = 8$,
 the achievable rate initially increases with the number of antennas but eventually decreases as the number of antennas continues to grow.
This is because when the number of antennas is small,  the scattering paths cannot be distinguished in angle domain accurately, leading to significant inter-path interference.            
As the number of antennas increases, the resolution in the angle domain improves, thereby enhancing the achievable rate.                     
However, when the number of antennas becomes excessively large, the beam squint effect becomes significant, resulting in a decrease in NMSE performance. 
 As a result, as the number of antennas increases, the achievable rate progressively approaches  the optimal value.
Traditional precoding cannot eliminate the effect of beam squint, resulting the achievable rate remains the lowest as the number of antennas increases.
}

\begin{figure}[htbp]
 \centering
 \includegraphics[width=75mm]{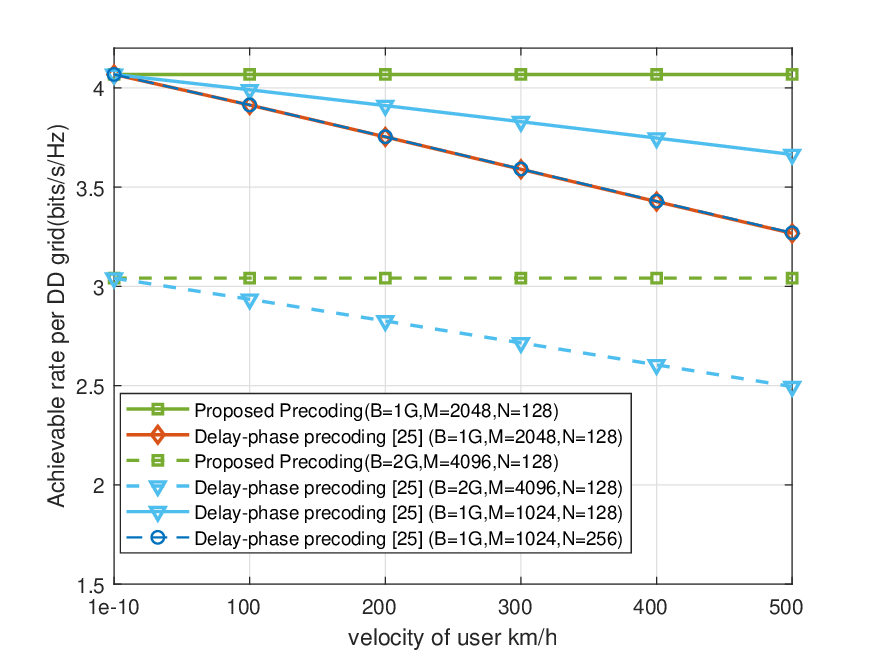}
 \caption{Achievable rate per DD grid versus the velocity of user.}
 \label{rate vs velocity}
\end{figure}

{
\figurename{ \ref{rate vs velocity}}  illustrates the achievable rate per DD grid versus the velocity of user, where the impact of bandwidth and subcarrier spacing on system performance are analyzed. 
From \figurename{ \ref{rate vs velocity}},
when the bandwidth increases, the beam squint effect increases, which leads to the decrease of the achievable rate.
On the other hand, when
% the bandwidth is constant and the subcarrier spacing is increased, the number of subcarriers decreases to 1024.
%In this case,
the bandwidth is fixed and the product $M\times N$ decreases, the achievable rate decreases.
This is due to the fact that decreases  in $M\times N$ leading to a
 reduction in symbol duration.
This reduction in symbol duration mitigates the Doppler squint effect from the analysis in Section II.D.
Furthermore, Fig. 12 shows that when $M\times N$ and velocity are constant, the impact of Doppler squint remains unchanged.
This phenomenon validates our previous analysis. 
}

\section{Conclusion}
%In this paper, we investigate Doppler Squint Effect and Beam Squint Effect in massive MIMO OTFS system.
%The effects of Doppler Squint and beam Squint on massive MIMO OTFS are analyzed.
%Specifically, the receiving beamforming for chirp sequence is illustrated and
%the channel parameters estimation based on chirp sequence was proposed. Then, a hybrid downlink OTFS precoding was proposed. The effects of beam Squint and Doppler Squint are reduced by using a hybrid precoding scheme.
%Finally, the performance of the proposed precoding is analyzed to verify the effectiveness of the proposed algorithm.

This paper { introduced}  a massive MIMO-OTFS system that accounts for the doubly squint effect, i.e., beam squint and Doppler squint effect.
The input-output relationship was derived, which  facilitated the analysis of doubly squint effect. 
Then, a channel estimation method with chirp pilot was proposed, where the parameters are determined by the peak index and peak value of DFT-angle sequences. 
For the receiving processing of the BS, a TTDs and PSs configuration was devised to mitigate the doubly squint effect. 
Additionally, a hybrid precoding scheme was introduced to compensate the phase shifts caused by doubly squint effect. 
Finally, simulation results substantiate the superiority and viability of the proposed methods.

{
It is worth noting that the channel model used in this work can still be regarded as a sparse model, where the number of scattering paths is finite and separable. 
However, the channel of practical scenarios may be non-sparse, leading to the spread of multipath components and the spread caused by doubly squint effect intertwined.
Thus, the performance can be deteriorates severely in practical scenarios due to channel mismatch.
Research on the doubly squint effect in non-sparse scenarios still requires further efforts to study.
}

% or


\appendices

\begin{figure*}
\begin{subequations}
{ 
\begin{align}
& {\tilde h}_{k,\ell}^{u,u'}[k' \!,\ell']   \!
= \!
{\frac{{\sqrt{\rho _{c}}}}{{MN}}}  
\!
\sum\limits_{n = 0}^{N-1}
\sum\limits_{m = 0}^{M-1}
 \sum\limits_{d = 1}^{N_T} \sum\limits_{s = 1}^{N_P}   \sum\limits_{p = 1}^{P} {\sum\limits_{n' = 0}^{N - 1} }
\sum_{c = 1}^{N_R} 
\! \int_{(n-n')T}^{(n-n'+1)T} 
\! \! \! \! \!
 {{\alpha }_{p,u}} \left| {{\mu _{p,u}}} \right| 
 {e^{\! -j2\pi {{\nu_{p\!,u}}\mu _{p,u}}\left(\tau \!  -t_{c,d} \! - \tau _{p,u} \! - \! (a-1)\frac{{{\psi _{p,u}}}}{{{f_c}}} \right) }}e^{ \! -j2\pi \tau m \Delta f } \notag \\
 & \times \!
e^{-j2\pi f_c \! \left( \! t_{c,d}+ (a \! - \! 1)\frac{\psi _{p,u}}{f_c} \right) } \!
 \int {{e^{j2\pi \nu\left({\mu _{p,u}}\left(\tau \! -t_{c,d}- \tau _{p,u} \! - (a-1)\frac{{{\psi _{p,u}}}}{{{f_c}}}\right) + n'T + \tau  + \frac{{i'T}}{M}\right)}}}  
   {e^{j2\pi (\frac{{n' \! k'}}{N} \! - \!\frac{{mi'}}{M})}}
{e^{ \! - j2\pi {\Psi _{\! c ,d ,s}}}} {e^{j2\pi \left( \! - \! \frac{{nk}}{N} \! +  \! \frac{{m\ell}}{M}\right)}}d\nu d\tau
\label{h ap1 a} \\ 
&= \! {\frac{{\sqrt{\rho _{c}}}}{{MN}}}  \!
\sum\limits_{n = 0}^{N-1}
\sum\limits_{m = 0}^{M-1}
 \sum\limits_{d = 1}^{N_T} \sum\limits_{s = 1}^{N_P}   \sum\limits_{p = 1}^{P} {\sum\limits_{n' = 0}^{N - 1} }
\sum_{c = 1}^{N_R} \int_{(n-n')T}^{(n-n'+1)T} 
\!\!\!\!\!
 {{\alpha }_{p,u}} \left| {{\mu _{p,u}}} \right| 
 {e^{\! -j2\pi {{\nu_{p,u}}\mu _{p,u}}\left(\tau \!-t_{c,d} \! - \tau _{p,u} \! - (a-1)\frac{{{\psi _{p,u}}}}{{{f_c}}}\right) }}e^{-j2\pi \tau m \Delta f }
 {e^{j2\pi \left(-\frac{{nk}}{N} + \frac{{m\ell}}{M}\right)}}
 \notag \\
&\times \! e^{\! -j2\pi f_c \left(t_{c,d} + (a\! -1)\frac{\psi _{p,u}}{f_c} \right) 
}  \delta \! \left( \! (1 \! + \! \mu _{p,u}\! )\tau  \! - \! \left( \!\mu _{p,u} \! \left( \!t_{c,d} \! + \! \tau _{p,u} \! + \! (  a \! - \! 1 \! )\frac{\psi_{p,u}}{f_c}\right)\! - \! n'T \! -  \!\frac{i'T}{M}   \right) \right) {e^{j2\pi \left(\frac{{n'\!k'}}{N} \! - \! \frac{{mi'}}{M}\right)}}
 {e^{ \! - j2\pi {\Psi _{a\! ,b \! ,c}}}} d\tau
\label{h ap1 b} \\ 
&\overset{n'=n}{=}
\sum\limits_{n = 0}^{N-1}
\sum\limits_{m = 0}^{M-1}
 \sum\limits_{d = 1}^{N_T} \sum\limits_{s = 1}^{N_P}  \sum\limits_{p = 1}^{P}\sum_{c = 1}^{N_R} 
\frac{{\sqrt{\rho _{c}}}}{{{NM} }}
{{\alpha }_{p,u}}\left| {\frac{{{\mu _{p,u}}}}{{{\mu _{p,u}} + 1}}} \right|{e^{ - j2\pi  m\Delta f\left(t_{c,d}+{\tau _{p,u}} + (a-1)\frac{{{\psi _{p,u}}}}{{{f_c}}}\right)}} e^{-j2\pi f_c \left({t_{c ,d}}+(a-1)\frac{\psi _{p,s}}{f_c} \right)}  \notag \\
&\times
{e^{j2\pi \left(\frac{{{\mu _{p,u}}}}{{{\mu _{p,u}} + 1}}{\nu_{p,u}} + \frac{{m\Delta f}}{{{\mu _{p,u}} + 1}}\right)\left(t_{c,d}+{\tau _{p,u}} + (a-1)\frac{{{\psi _{p,u}}}}{{{f_c}}} + nT + \frac{{i'T}}{M}\right)}}{e^{j2\pi \left(\frac{{nk'}}{N} - \frac{{mi'}}{M}\right)}}
%\notag \\
%&\times
{e^{ - j2\pi {\Psi _{c ,d ,s}}}} 
{e^{j2\pi \left(-\frac{{nk}}{N} + \frac{{m\ell}}{M}\right)}}
\label{h ap1 c} \\ 
&  \overset{i'=l'}{\approx}  
\sum\limits_{n = 0}^{N-1}
\sum\limits_{m = 0}^{M-1}
 \sum\limits_{p = 1}^{P}
\sum\limits_{c = 1}^{N_R} 
\sum\limits_{d = 1}^{N_T} 
\sum\limits_{s = 1}^{N_P} 
{\frac{{\sqrt{\rho _{c}}}}{ {NM} }}
{{\alpha }_{p,u}}
  e^{j2\pi\left(\frac{n(k'-k)}{N}- \frac{m(\ell'-\ell)}{M}\right) } 
 {e^{ - j2\pi ( {{ t_{c,d}}{f_c}} + {((d-1)N_P+s-1){\psi _{p,u}}})\left(1 + \frac{{m\Delta f}}{{{f_c}}}\right)}}
\notag \\
&\times
 {e^{  j2\pi {\Psi _{c ,d ,s}}}}
 {e^{j2\pi {\nu_{p,u}}\frac{{\left({   \ell'+\ell_{p,u}+ {\tilde \ell}_{c,d} + ((d-1)N_P +s -1)\frac{\psi_{p,u}}{f_cT_s}  } \right)T}}{M}}}
{e^{ - j2\pi \frac{{m{\ell_{p,u}}}}{M}}}
{ e^{j2\pi n\left(\frac{{{k_{p,u}}}}{N} + \frac{m}{{{\mu _{p,u}}}}\right)}} 
\label{h ap1 d}
\end{align} 
}
% \hrulefill
\end{subequations}
\end{figure*}

\section{Proof of DD-Domain Equivalent Channel Response without Involving the Digital Precoding}
\label{proof of huu}
We first consider the case of ICI (i.e., $n^{\prime} = n$). Given $B_{c}[n, m] = 1$ and $D_{c, u}[k, \ell] = 1, \forall c,u,m, n, k, \ell$, the DD-domain equivalent channel response can be represented as $\tilde{h}_{k,\ell}^{u, u^{\prime}}\left[k^{\prime},\ell^{\prime}\right]$, given by \eqref{h ap1 d}. Here, \eqref{h ap1 b} is derived because the summation 
$ \sum_{m^{\prime}=0}^{M-1} e^{j 2\pi \frac{m^{\prime} (i - \ell^{\prime})}{M}}$
is non-zero exclusively when $i - \ell^{\prime} = 0$. 
From \eqref{h ap1 b}, we can derive that $\tilde{h}_{k,\ell}^{u, u^{\prime}}\left[k^{\prime},\ell^{\prime}\right]$ is non-zero when $\tau$ is derived by
\begin{align} \label{tau eq}
 (1\! +\! \mu_{p, u}) \tau \! = \!  \mu_{p, u}  \! \left( \tau_{p, u}\! + \! t_{c,d} \! + \! (s\! -\! 1)\! \frac{\psi_{p, u}}{f_{c}}\! \right)\! -  \!n^{\prime} T \! - \!\frac{i^{\prime} T}{M} \!, \!
\end{align}
where $i^{\prime} = \ell^{\prime}$. 
Considering $(n-n')T  <\tau < (n-n'+1)T $, equality \eqref{tau eq} is valid only when we have
\begin{align} \label{idx set ap1 ici}
0  \! \le  \! \ell'  \! <  \! M   \! \! -  \! \left(  \!{ \ell_{p',u'}}  \! \!+ \! \ell_{c,d} \! +  \! (  \!a  \! -  \! 1  \! )  \! \frac{{{\psi _{p',u'}}M}}{{{f_c}T}}\right)  \! \! +  \! \frac{{M(n  \! +  \! 1)}}{{{\mu _{p',u'}}}}.
\end{align}
% \begin{figure*}
% \begin{align}\label{u l 0}
% \left\{ \begin{array}{l}
% (1 + {\mu _{p,u}})(n - n')T - {\mu _{p,u}}\left({\tau _{p,u}} + (a-1)\frac{{{\psi _{p,u}}}}{{{f_c}}}\right) + n'T + \frac{{i'T}}{M} < 0\\
% (1 + {\mu _{p,u}})\left ((n - n' + 1)T - \frac{{i'T}}{M} \right ) - {\mu _{p,u}} \left({\tau _{p,u}} + (a-1)\frac{{{\psi _{p,u}}}}{{{f_c}}} \right) + n'T + \frac{{i'T}}{M} > 0
% \end{array} \right. ,
% \end{align}
% \begin{align}\label{u s 0}
% \left\{ \begin{array}{l}
% (1 + {\mu _{p,u}})(n - n')T - {\mu _{p,u}}\left ({\tau _{p,u}} + (a-1)\frac{{{\psi _{p,u}}}}{{{f_c}}}\right ) + n'T + \frac{{i'T}}{M} > 0\\
% (1 + {\mu _{p,u}})\left((n - n' + 1)T - \frac{{i'T}}{M}\right) - {\mu _{p,u}}\left ({\tau _{p,u}} + (a-1)\frac{{{\psi _{p,u}}}}{{{f_c}}}\right) + n'T + \frac{{i'T}}{M} < 0
% \end{array} \right. ,
% \end{align}
% \hrulefill
% \end{figure*}
% We denote $\mathcal{L}_{ICI}^{p, u}$ as the index set of ICI with respect to the $p$-th path with $u$-th user. 
Then, the index set of ICI $\mathcal{L}_{ICI}^{p, u}$ can be derived by substituting $i^{\prime} = \ell^{\prime}$ into \eqref{idx set ap1 ici}.
 The approximation in \eqref{h ap1 d} arises because $\mu_{p, u} \approx \mu_{p, u} + 1$.
 % which leads to the approximation $\mu_{p, u} \approx \mu_{p, u} + 1$.

The analysis of the ISI term is similar except for two differences. The first on is the integration range in equation \eqref{h ap1 b} extending from \((n - n' - 1)T\) to \((n - n')T\). The second one is to employ $n' = [n - 1]_N$ in equation \eqref{h ap1 c}. With these considerations, we can derive the ISI index set for \(\ell'\) as \(\mathcal{L}_{ISI}^{p,u}\). Consequently, the DD-domain equivalent channel response without involving the digital precoding can be represented by equation \eqref{H_p_l from dd domain to tf domain}.

\section{Proof of Proposed Hybrid Precoding Method}
\label{proof of bd after precoding}

% \begin{figure*}[ht]
% \begin{align}\label{proof for index set}
%   h_{k,\ell}^{u,u'}[k',\ell'] 
% =& \frac{{\hat {\bar \alpha} _{p,u}^*}}{{\left| {{{\hat {\bar \alpha} }_{p,u}}} \right|}}
% {\frac{1}{{MN}}}  
% \sum\limits_{n = 0}^{N-1}
% \sum\limits_{m = 0}^{M-1}
%  \sum\limits_{d = 1}^{N_T} \sum\limits_{s = 1}^{N_P}   \sum\limits_{p = 1}^{P} {\sum\limits_{n' = 0}^{N - 1} }
% \int_{(n-n')T}^{(n-n'+1)T} 
%  {{\alpha }_{p,u}} \left| {{\mu _{p,u}}} \right| {e^{-j2\pi {\mu _{p,u}}\left(\tau - \tau _{p,u} - (s-1)\frac{{{\psi _{p,u}}}}{{{f_c}}}\right) {\nu_{p,u}}}}e^{-j2\pi \tau m \Delta f } \notag \\
% &\times \delta\left((1+\mu _{p,u})\tau - \left(\mu _{p,u} \left(\tau _{p,u}+(c-1)\frac{\psi_{p,u}}{f_c} \right)-n'T- \frac{i'T}{M}   \right)\right)d\tau {e^{j2\pi \left(\frac{{n'k'}}{N} - \frac{{mi'}}{M}\right)}}{e^{ - j2\pi n'\frac{{{{\hat k}_{p,u}}}}{N}}} {e^{ - j2\pi {{\hat \nu}_{p,u}}\frac{{\ell'T}}{M}}}
% \notag \\
% &\times {e^{j2\pi \left(-\frac{{nk}}{N} + \frac{{m\ell}}{M}\right)}}
% \end{align} 
% \hrulefill
% \end{figure*}

\begin{figure*}[ht]
\vspace{-0.8cm}
\begin{subequations}
  { 
    \begin{align}
 & h_{k,\ell}^{u',u'}[k',\ell'] 
\overset{}{=} 
\frac{\hat \alpha_{p,u} ^*}{\left| \hat\alpha_{p,u}  
 \right|}
{\frac{N_T}{{MN}}}  
\sum\limits_{n = 0}^{N-1}
\sum\limits_{m = 0}^{M-1}
 \sum\limits_{s = 1}^{N_P}   \sum\limits_{p'= 1}^{P'} {\sum\limits_{n' = 0}^{N - 1} }
\int_{(n-n')T}^{(n-n'+1)T}  \! \! \!
 {{\alpha }_{p',u'}} \left| {{\mu _{p',u'}}} \right| {e^{-j2\pi {\mu _{p',u'}}\left(\tau \! - \! \tau _{p',u'}\! -\!(s-1)\frac{{{\psi _{p',u'}}}}{{{f_c}}}\right) {\nu_{p',u'}}}}e^{-j2\pi \tau m \Delta f }
\notag \\
 & \! \times \! \int {{e^{j2\pi \nu \left({\mu _{p',u'}} \left(\tau - \tau _{p',u'} - (s-1)\frac{{{\psi _{p',u}}}}{{{f_c}}}\right) + n'T + \tau  + \frac{{i'T}}{M}\right)}}} 
   {e^{j2\pi \left(\frac{{n'k'}}{N} \! - \! \frac{{mi'}}{M}\right)}} {e^{ - j2\pi n'\frac{{{{\hat k}_{p',u'}}}}{N}}} {e^{ \! - j2\pi {{\hat \nu}_{p',u}}\frac{{(\ell' \! +  \bar \ell_{u'})T}}{M}}}
{e^{j2\pi \left(\! -\frac{{nk}}{N} + \frac{{ml}}{M}\right)}}  d\nu d\tau
\label{huua}
\\ 
 &\approx 
{{\left| {{{ \alpha }_{p',u'}}} \right|}} 
\sum\limits_{n = 0}^{N-1}
\sum\limits_{m = 0}^{M-1}
  \sum\limits_{p' = 1}^{P'}
e^{-j2\pi \frac{m(\ell'-\ell-{\bar \ell}_{u'})}{M}}
e^{j2\pi \frac{n(k'-k)}{N}} \label{huud}
% \sum\limits_{c = 1}^{N_P} e^{-j2\pi (s-1)\psi_{p',u'}\frac{m\Delta f-\nu_{p',u'}}{f_c} }
\end{align} 
}
\end{subequations}
\hrulefill
\end{figure*}

We first consider the case of ICI.
By substituting \eqref{ttd in precoding}, \eqref{ps in precoding}, \eqref{DD domain precoding} and \eqref{BTF} into \eqref{H_simple},
%We first consider the case of ICI.
%By combining \eqref{DD comp}, \eqref{TF comp}, \eqref{h_tau_v2}, \eqref{Hp}, 
% %\eqref{ysnm}, \eqref{hkl}, 
%\eqref{recv Y including ICI and ISI}, 
%%\eqref{recv y dd ISI ICI},
% \eqref{ttd in precoding}, \eqref{ps in precoding}, \eqref{DD domain precoding} and \eqref{BTF},
we can derive \eqref{huua},
% at the top of the next page.
 where, $i'= \ell' - \hat{\ell}_{p',u'} - \bar{\ell}_{u'} + \frac{Mn'}{\hat{\mu}_{p',u'}}$. 
 Similar to the approach in Appendix A,
%  \eqref{huuc} can be derived when $(1+\mu _{p',u'})\tau - \left(\mu _{p',u'} \left(\tau _{p',u'}+(s-1)\frac{\psi_{p',u'}}{f_c} \right)-n'T- \frac{i'T}{M} \right)= 0$.
% Similar to \eqref{u l 0} and \eqref{u s 0}, we can derive the the effective range of $i'$ for $\mu_{p,u}>0$ and $\mu_{p,u}<0$, respectively.
% Then, 
% we can derive \eqref{huua} when
the index set for $i'$ is given by
\begin{align}\label{ICI for i}
0 \le i' \! < \! M  \! - \left({ \ell_{p',u'}} + (s \! - \! 1 )\frac{{{\psi _{p',u'}}M}}{{{f_c}T}}\right) \! + \! \frac{{M(n + 1)}}{{{\mu _{p',u'}}}}.
\end{align}
Considering $i'= \ell' - \hat{\ell}_{p',u'} - \bar{\ell}_{u'} + \frac{Mn'}{\hat{\mu}_{p',u'}}$, the index set for $l'$ can be represented as
\begin{align}\label{ICI for l}
\bar{\ell}_{u'} + {\hat \ell_{p',u'}} \! - \!\frac{{Mn'}}{{{\mu _{p',u'}}}} \le \ell' < M \! - \! (s \! - \! 1) \frac{{{\psi _{p',u'}}}}{{{f_c}{T_s}}} + \frac{M}{{{\mu _{p',u'}}}} + \bar{\ell}_{u'} .
\end{align}
Similarly, we can derive index set for ISI as
\begin{align}\label{ISI for l}
(s \! - \!1) \frac{{{\psi _{p',u'}}}}{{{f_c}{T_s}}} \! + \! \frac{M}{{{\mu _{p',u'}}}} \! + \! \bar{\ell}_{u'}
\le \ell' <  \bar{\ell}_{u'} + {\hat \ell_{p',u'}} \! - \! \frac{{Mn'}}{{{\mu _{p',u'}}}}.
\end{align}
From \eqref{ICI for l} and \eqref{ISI for l}, the index set for ICI and ISI is related to the PS index $s$.
Since each RF chain connects $N_P$ shifters, we expect the index set to be suitable for all phase shifters.
With $\ell'$ being an integer in OTFS modulation, \eqref{ICI for l} can be simplified as
\begin{align}\label{ICI index set afp}
&\mathcal{L}_{ICI}^{\;p',u'} = 
%\notag \\
\left \{ \ell'\in \mathbb{N}:  { \hat \ell_{p',u'}} 
 %+ \frac{{{{\tilde t}_{p,i}}}}{{{T_s}}} 
+\bar {\ell}_{u'} - 1
% \frac{{Mn'}}{{{\hat \mu _{p',u'}}}} 
\leqslant \ell' \leqslant M - 1 \right \},
%\frac{{{c \hat \psi _{p,i}}}}{{{f_c}{T_s}}} + \frac{M}{{{\mu _{p,i}}}}  + \bar {l}_{p,s}\} ,
%  \end{align}
%  \begin{align}
%\label{ISI boundry}
\end{align}
%\end{figure*}
where we have
\begin{subequations}
   \begin{align}\label{st}
& 0 < -  \frac{{{(s-1) \hat \psi _{p',u'}}}}{{{f_c}{T_s}}} + \frac{M}{{{\hat \mu _{p',u'}}}} + \bar {\ell}_{u'} < 1,
\end{align}
\begin{align}
 \label{st2}
 &0 < \bar {\ell}_{u'} -  \frac{Mn'}{{{\hat \mu _{p',u'}}}} < 1,\quad 
\end{align} 
\end{subequations}
$\forall$ $1\le s \le N_P$, $0\le n' \le N-1$ and $ \left | \frac{1}{\hat \mu_{p',u'}} \right| < \frac{\nu_{max}}{f_c} $. Here, $\nu_{max}$ is the maximum Doppler shift. 
We omit the analysis of ISI due to the space limitation. The index set for ISI can be expressed as
\begin{align}\label{ISI index set afp}
&\mathcal{L}_{ISI}^{\;p',u'} =
%\notag \\& 
\left \{\ell'\in \mathbb{N}:  
%-  \frac{{{c \hat \psi _{p,s}}}}{{{f_c}{T_s}}} + \frac{M}{{{\hat \mu _{p,s}}}} + \bar {l}_{p,s}
1 \leqslant \ell' < { \hat \ell_{p',u'}} + \bar {\ell}_{u'}
%+ \frac{{{{\tilde t}_{p,i}}}}{{{T_s}}} 
- 1
% \frac{{Mn'}}{{{\hat \mu _{p',u'}}}} 
\right \}.
\end{align}
After that, we can get \eqref{huud}, which does not contain path-related terms. That is, our precoding can compensate the phase deviation introduced by different paths. 
Considering $\bar{\ell}_{u'}$ is constant, the phase deviation introduce by $\bar{\ell}_{u'}$ can be easily compensated at receiver.


\end{document}